\pdfoutput=1 
\documentclass[%
reprint,
superscriptaddress,
 amsmath,amssymb,
 aip,
]{revtex4-1}

\usepackage{graphicx}% Include figure files
\usepackage{dcolumn}% Align table columns on decimal point
\usepackage{bm}% bold math
\usepackage[labelfont=bf,labelsep=period,justification=raggedright]{caption}
\usepackage{setspace}

\begin{document}
%\preprint{APS/123-QED}

\title{Functional complexity emerging from anatomical constraints in the brain: the significance of network modularity and rich-clubs.}

\author{Gorka Zamora-L\'opez} \email{gorka@Zamora-Lopez.xyz}
\affiliation{Center for Brain and Cognition, Universitat Pompeu Fabra, Barcelona, Spain}
\affiliation{Department of Information and Communication Technologies, Universitat Pompeu Fabra, Barcelona, Spain}
\author{Yuhan Chen}
\affiliation{Department of Physics, Hong Kong Baptist University, Hong Kong, P.R. China}
\affiliation{Centre for Nonlinear Studies, Hong Kong Baptist University, Hong Kong, P.R. China}
\affiliation{State Key Laboratory of Cognitive Neuroscience and Learning, Beijing Normal University, P.R. China}
\author{Gustavo Deco}
\affiliation{Center for Brain and Cognition, Universitat Pompeu Fabra, Barcelona, Spain}
\affiliation{Department of Information and Communication Technologies, Universitat Pompeu Fabra, Barcelona, Spain}
\affiliation{Instituci\'o Catalana de la Recerca i Estudis Avan\c cats, Universitat Pompeu Fabra, Barcelona, Spain}
\author{Morten L. Kringelbach}
\affiliation{Department of Psychiatry, University of Oxford, Oxford, UK}
\affiliation{Center of Functionally Integrative Neuroscience (CFIN), Aarhus University, Aarhus, Denmark}
\affiliation{Oxford Functional Neurosurgery and Experimental Neurology Group, Nuffield Departments of Clinical Neuroscience and Surgical Sciences, University of Oxford, UK}
\author{Changsong Zhou}
\affiliation{Department of Physics, Hong Kong Baptist University, Hong Kong, P.R. China}
\affiliation{Centre for Nonlinear Studies, Hong Kong Baptist University, Hong Kong, P.R. China}
\affiliation{The Beijing-Hong Kong-Singapore Joint Centre for Nonlinear and Complex Systems, Hong Kong, P.R. China}
\affiliation{Beijing Computational Science Research Center, Beijing, China}
\affiliation{Research Centre, HKBU Institute of Research and Continuing Education, Shenzhen, China}

\begin{abstract}
The large-scale structural ingredients of the brain and neural connectomes have been identified in recent years. These are, similar to the features found in many other real networks: the arrangement of brain regions into modules and the presence of highly connected regions (hubs) forming rich-clubs. Here, we examine how modules and hubs shape the collective dynamics on networks and we find that both ingredients lead to the emergence of complex dynamics. Comparing the connectomes of \emph{C. elegans}, cats, macaques and humans to surrogate networks in which either modules or hubs are destroyed, we find that functional complexity always decreases in the perturbed networks. A comparison between simulated and empirically obtained resting-state functional connectivity indicates that the human brain, at rest, lies in a dynamical state that reflects the largest complexity its anatomical connectome can host. Last, we generalise the topology of neural connectomes into a new hierarchical network model that successfully combines modular organisation with rich-club forming hubs. This is achieved by centralising the cross-modular connections through a preferential attachment rule. Our network model hosts more complex dynamics than other hierarchical models widely used as benchmarks.
\end{abstract}

\maketitle

%#########################################################################################
\section{Introduction}

The study of interconnected natural systems as complex networks has uncovered common principles of organisation across scientific domains. Two pervasive features are ($i$) the grouping of the nodes into modules and ($ii$) the presence of highly connected nodes or hubs. It was soon recognised that these two features are signatures of hierarchical organisation but attempts to incorporate both into realistic network models have been of limited success.~\cite{Ravasz_Hierarchical_2003}. Currently, the most popular hierarchical models recursively divide modules into smaller modules~\cite{Arenas_SynchScales_2006}. These networks, however, lack of hubs. Investigation of the brain's connectivity has shed light on how nature efficiently combines the two features. Real connectomes are modular with the cross-modular connections centralised through highly connected brain regions which form a rich-club~\cite{Zamora_Hubs_2010, Heuvel_HubsHuman_2011, Varshney_Elegans_2011}.

The nervous system acquires information about the environment through different channels, known as sensory modalities. Information from each channel is independently processed by specialised neural compartments. An adequate and efficient integration of the information of those different channels is necessary for survival~\cite{Damasio_BrainBinds_1989, FusterBook_2003}. In a series of numerical experiments, Tononi and Sporns attempted to identify the right topologies that help optimally balance the coexistence of both segregated subsystems and an efficient integration of their information.~\cite{Sporns_Classes_2001}. Starting from an ensemble of random graphs, an evolutionary algorithm would select those networks with the largest complexity. In subsequent iterations the winners would be mutated  -- slightly rewired -- to produce another population to start over. The underlying assumption was that an increase of the \emph{neural complexity} defined by the authors would lead to networks with balanced capacity to integrate and segregate information~\cite{Tononi_Complexity_1994}. This procedure gave rise to networks with interconnected communities capturing the relevance of modules for the segregation of information. However, the optimised networks lacked of hubs and rich-clubs. Dynamical models on modular networks have shown that there is a balanced rate in the number of inter- to intra-modular links that optimises the complexity of the network dynamics~\cite{Zhao_Competition_2011, Adjari_Crossover_2012}. This phenomenon has also been observed in contagion spreading, where the contagion threshold depends on the node's degree~\cite{Nematzadeh_Contagion_2014}. Too few links between the communities leads to clustered (segregated) dynamics but no efficient interaction between them. On the contrary, too many connections between communities easily leads to a globally synchronised network meaning there is integration but no dynamical segregation. A balance is achieved in between. However, 
it can be argued that in modular networks integration is not efficient because it happens via global synchrony, which is an undesirable state of neural networks. 

Despite these and other past efforts, the relation between a network's complexity and its capacity to segregate and integrate information is yet unresolved and confusing. In particular, their causal relation requires clarification. While it seems plausible to assume that the needs of neural systems to integrate and segregate information may have led to the development of complex topological features, e.g., modules and rich-clubs, the opposite is not necessarily true. A network optimised for high complexity does not necessarily end developing modules and hubs, nor being good for integrating and segregating information. The aim of the present paper is to test and corroborate this causal relation only in one direction, namely, that the hierarchical centralisation of cross-modular connections through rich-clubs leads to enhanced functional complexity. For that purpose we consider both real neural connectomes and synthetic network models. We study the evolution of their functional complexity as the networks undergo a transition towards global synchrony by gradually increasing the weights of the links. We find that functional complexity emerges for intermediate values of the tuning parameter; when the nodes are neither independent from each other nor globally synchronised.

By comparing the real networks to randomised versions in which either the presence of hubs or the modular structure are destroyed, we find that both topological features are crucial ingredients for the networks to achieve high functional complexity. In the randomised networks complexity is always reduced. To clarify the precise impact of rich-clubs we have also carried out a lesion study. Selective removal of the links between the hubs leads to a reduction of functional complexity in all cases. The reduction is significant compared to random lesions. In the case of the human dataset we also observe that the dynamics of the brain, at rest, reflects a state with the largest complexity that its anatomical connectome can host. Last, we introduce a new model of hierarchical networks inspired on the topology of neural and brain networks. Our hierarchical network model successfully combines nested modules with the presence of hubs. This is achieved by centralising the inter-modular connectivity through a few nodes by a preferential attachment rule. These networks achieve higher complexity than other well-known benchmark models.

The manuscript is organised as follows. First we introduce a measure of functional complexity that is based on the variability of the pair-wise cross-correlations of the nodes. We then investigate the complexity of neural networks in comparison to surrogate networks. Finally we compare the complexity of common random and hierarchical network models and we introduce the new model of modular and hierarchical networks with centralised inter-modular connectivity.

%#########################################################################################
\section{Measuring functional complexity}

Despite the common use of the term ``complex networks'' a formal quantitative measure is missing to determine how complex a network is. Here we take an indirect approach and estimate the complexity of the collective dynamics that the network can host. In general, the complexity of a coupled dynamical system is a combination of the temporal complexity of the signals traced by the individual nodes and of the spatial formation of clusters. Because we are here interested on the influence of the network's topology and because the temporal complexity depends on the model chosen for the node dynamics, we study the spatial aspect of complexity. We refer to this as \emph{functional complexity} for consistency with the term \emph{functional connectivity} to denote the time-averaged dynamical interdependencies between neural populations.

\begin{figure}[!ht]
\begin{center}
\includegraphics[width=3.27in]{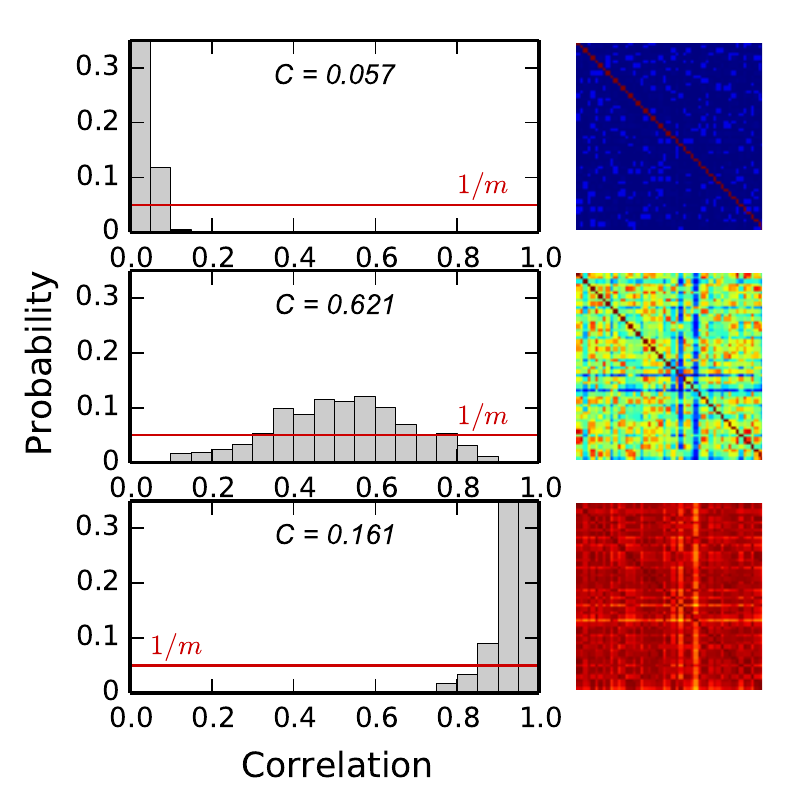}
\end{center}
\caption{{\bf Illustration of the measure for functional complexity.}
When the collective dynamics of a network are close to independence or to global synchrony (top and bottom panels) the distribution of the cross-correlation values are characterised by narrow peaks close to $r_{ij} = 0$ or to $r_{ij} = 1$. Complex dynamical interactions happen when the collective behaviour is characterised by intermediate states leading to a broad distribution of the correlation values (middle panel). Red lines correspond to the uniform distribution, $\bar{p} = \frac{1}{m}$, where $m$ is the number of bins.
} %end of caption
\label{fig:Complexity}
\end{figure}

Given a network of $N$ coupled dynamical nodes, e.g. neurones, cortical regions or oscillators, its pair-wise correlation matrix $R$ reflects the degree of interdependencies among the nodes. When the nodes are disconnected from each other, they are also dynamically independent and hence, no complex collective dynamics emerge. All correlation values are $r_{ij} \approx 0$, Fig.~\ref{fig:Complexity}(top). In the opposite extreme, when the nodes are strongly coupled, the network becomes synchronised. However, global synchrony is neither a complex state because all nodes follow the same behaviour. In this case all the correlation values are $r_{ij} \approx 1$, Fig.~\ref{fig:Complexity}(bottom). Complexity emerges when the collective dynamics are characterised by intermediate states, between independence and global synchrony. Such states are reflected by a broad distribution of $r_{ij}$ values, Fig.~\ref{fig:Complexity}(middle). 

At the two extreme cases, independence and global synchrony, the distribution $p(r_{ij})$ of pair-wise correlations is characterised by a narrow distribution. In between, at the range in which the network dynamics are more complex, the distribution becomes broader. After these observations we define the \emph{functional complexity} $C$ of the network as the variability of $p(r_{ij})$. Now, there are different manners to evaluate the variance of a distribution. For example, in Ref.~\cite{Zhao_Complexity_2010}, complexity was defined as the normed entropy of $p(r_{ij})$. Here, we choose to define complexity as the difference between the observed distribution $p(r_{ij})$ and the uniform distribution. If $p(r_{ij})$ is estimated in $m$ bins, the uniform distribution is $\bar{p}_\mu = \frac{1}{m}$ for all bins $\mu = 1, 2, \ldots, m$. Hence, functional complexity is quantified as the integral between the two distributions, which is replaced by the sum of their differences over the bins:
%%%%%%%%
\begin{equation}
C = \; 1 \, - \, \frac{1}{C_m} \: \sum_{\mu=1}^m \left| \, p_\mu(r_{ij}) - \frac{1}{m}  \right|, 
\label{eq:Complexity}
\end{equation}
%%%%%%%
where $| \cdot |$ means the absolute value and $C_m = 2 \, \frac{m-1}{m}$ is a normalisation factor that represents the extreme case in which the $p(r_{ij})$ is a Dirac-delta function $\delta_m$. That is, when all $r_{ij}$ values fall in the same bin as it happens when the nodes are either mutually independent or globally synchronised. Because we are only interested in the pair-wise interactions we discard the diagonal entries $r_{ii}$ from the calculations.

After comparing different alternatives to quantify $C$ we found that the measure in Eq.~(\ref{eq:Complexity}) to be the most convenient solution; see Supplementary Information. This choice turned to be the most sensitive to discriminate between network topologies and also the most robust to variation in the number of bins. The reason is that the integral does not simply evaluate the broadness of the distribution but, more generally, its divergence from uniformity. This measure of functional complexity is easy to apply to empirical and to simulated data. While in this paper we study cross-correlations, the measure can be applied to any other metric of pair-wise functional connectivity, e.g. mutual information.

%########################################################################################
\section{Functional complexity of neural connectomes}

In this section we investigate the functional complexity of anatomical brain and neural connectomes. We study the binary corticocortical connectivities of cats, macaque monkeys and humans, and also the neuronal wiring of the \emph{C. elegans} (see Materials and Methods). We will refer to these as the structural connectivities (SC) and we will denote their corresponding correlation matrices $R$ as their functional connectivities (FC). For each SC network we study the evolution of its FC as the collective dynamics undergo a transition from independence to global synchrony. This transition is controlled by increasing the weights, or coupling strength $g$, of the SC links. We compare the results to two types of surrogate networks: ($i$) rewired networks that conserve the degree distribution and ($ii$) random modular networks which preserve the community structure of the original network, see Materials and Methods. In the rewired networks the hubs are still present although the modular structure vanishes. The modularity preserving random networks conserve the number of links within and across modules but alter the degree distribution and the hubs disappear. For completeness, we also compare the results to those of random graphs with the same size and number of links. All results for surrogate networks are averages over $1000$ realisations. In order to quantify more precisely the impact of the rich-club, we also include a lesion study. After selective removal of the links between the rich-club hubs functional complexity is reduced. This reduction is compared to ensembles of randomly lesioned networks. 

\begin{figure*}[!ht]
\begin{center}
\includegraphics[width=6.83in]{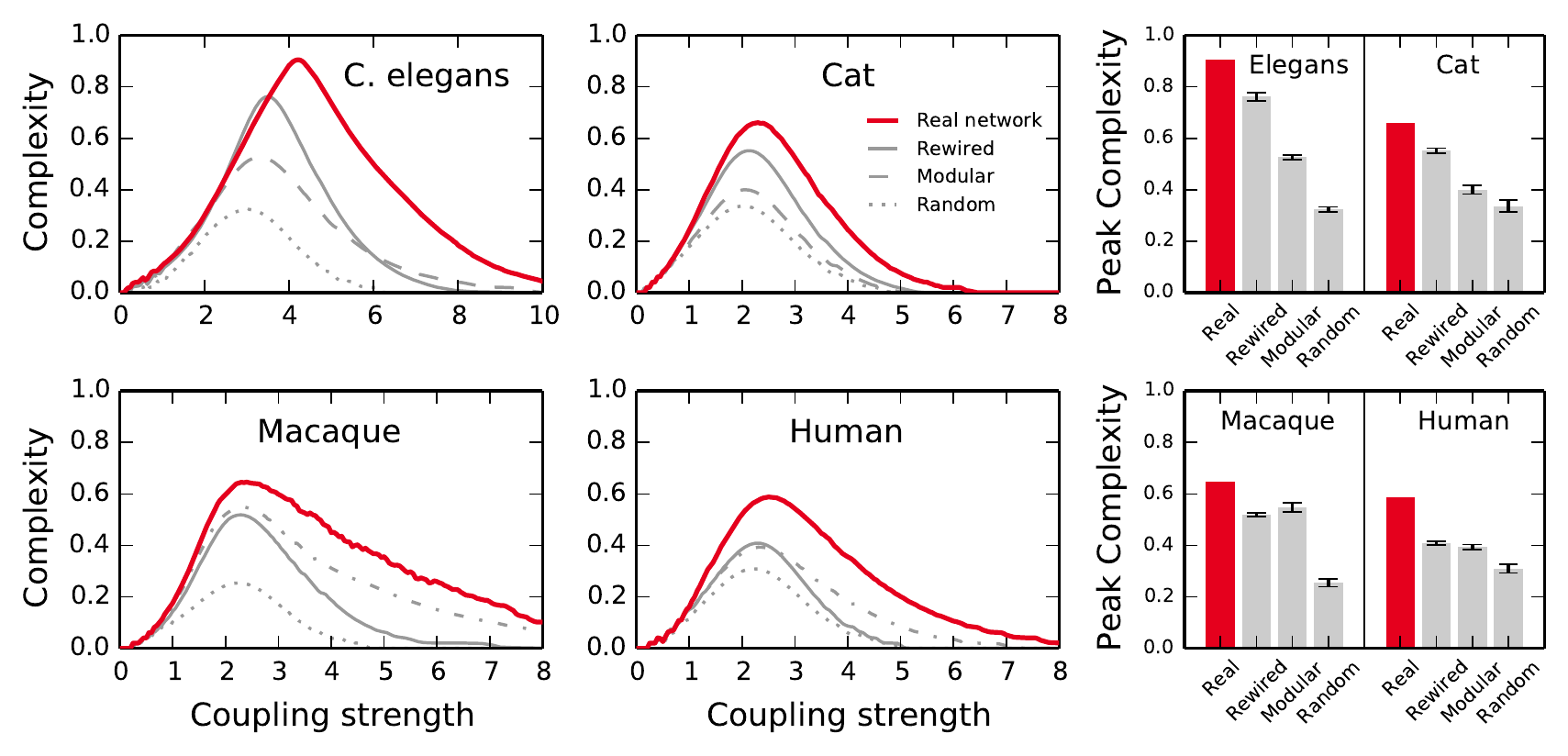}
\end{center}
\caption{
{\bf Functional complexity of anatomical connectomes.}
Comparison between the evolution of complexity for four neural networks (solid red lines) and the results for surrogate networks: random graphs (dotted lines), rewired networks conserving degrees of nodes (dashed lines), and modularity preserving random graphs (dashed lines). The right-hand panels summarise the peak complexities achieved in each case.
Note that absolute values are not comparable across species due to the different size and densities of the connectomes. Meaningful are the relative differences with the surrogates.
}%end of caption
\label{fig:RealNets}
\end{figure*}

To evaluate the functional complexity of the SC matrices we first need to estimate their FC matrices at different values of the coupling strength, $g$. Because we want to emphasise the contribution of the network's topology on the dynamics it can host we need to discard, as much as possible, other sources of influence on the network dynamics. For this reason we introduce a heuristic mapping to analytically estimate the correlation matrices $R$ out of the $SC$ without the need to run detailed simulations, Eqs.~(\ref{eq:COV}) and~(\ref{eq:Qexp}). See details in Materials and Methods. Assuming the network consists of a set of coupled Gaussian noise sources the time-averaged cross-correlation matrix $R$ of the system can be analytically estimated out of the structural connectivity matrix~\cite{Tononi_Complexity_1994}. In this framework the correlation between brain regions can be understood as the total \emph{influence} exerted by one region over another, accumulated over all possible paths of all lengths within the network. The coupling (the weight of the links), serves as a resolution parameter determining the range of correlations. When $g$ is weak perturbations quickly decay allowing only for local correlations between neighbouring nodes. As the coupling grows the range of the correlations gradually increases. For strong coupling, perturbations propagate along the whole network causing global correlations. An unrealistic property of the Gaussian diffusion model is that the system leads to divergent dynamics at strong couplings. Motivated by the fact that in neural systems the signals attenuate, that is, information fed into the network rapidly disappears or is transformed instead of perpetually propagate along the network, we solve the divergence problem introducing an exponential decay for the diffusion of the signals over longer paths. This exponential decay guarantees that, once the network is globally correlated, an increase in coupling has no influence and the system does not diverge. This property is shared by widely applied models for generic oscillatory and neural dynamics, e.g., Kuramoto oscillators and neural-mass models. Simulations performed with those models show the same qualitative behaviour as our exponential mapping; see Supplementary Information.

%________________________________________________________________________________________
\subsection{Comparison to surrogate networks}

The results for the neural and brain connectomes are shown in Fig.~\ref{fig:RealNets}. As expected, $C$ vanishes at the extremes, when $g = 0$ and when $g$ is large enough for the networks to globally synchronise. Complexity emerges at intermediate levels of $g$. Find sample correlation matrices in Supplementary Figure~S1. All real networks (solid red lines) achieve larger complexity than the surrogates along the whole range of $g$. The bar plots summarise the peak values. The lowest peak corresponds always to the random graphs (dotted lines) while the rewired (solid gray lines) and the modularity preserving (dashed lines) networks take intermediate complexities. These results show that it is the combination of hubs and modular structure what allows the real networks to reach larger functional complexities. Destroying one of these features, either the hubs (by randomising the networks to conserve only their modularity) or the modular structure (by rewiring links to conserve only the degrees), leads to a notable reduction in complexity. Another observation is that the transition to synchrony of the real networks is slower than that of the surrogates. This shows that there is a wide range of $g$ for which the complexity remains high.

Since the rewiring procedure does not necessarily disconnect the hubs from each other, it remains unclear what is the precise impact of the rich-club itself on the complexity. How is complexity altered when the hubs are disconnected from each other? In order to investigate this in more detail we have performed a lesioning study. First, we have identified the rich-clubs on each of the four empirical networks, see Supplementary Information, and then we have selectively removed all the links between the rich-club hubs. These comprise only a small fraction of the total number of links and thus small, but measurable changes in complexity are expected. After selective removal of the rich-club links from the SC matrices, their corresponding FC matrices were computed for the optimal $g$ at which the complexity $C(real)$ of the original networks were maximal. We find that, compared to the original networks, the functional complexity $C(lesion)$ in the lesioned networks decreases in all the four cases. See Table~\ref{tab:Lesions}. 

\begin{table}
{\centering
\begin{tabular}{p{1.8cm} | p{1.3cm} p{1.5cm} p{2cm} | p{1.2cm}}
{\bf Network} 			& ${\bf C(real)}$ & ${\bf C(lesion)}$ & {\bf Difference}  & {\bf Prob.}  \\
\hline
\emph{C. elegans}	& 0.905  & 0.884  & - 2.32 \% 	& 0.0   \\ 
Cat 				& 0.658  & 0.641  & - 2.60 \% 	& 0.015  \\
Macaque			& 0.646  & 0.615  & - 4.80 \%	& 0.0  \\
Human			& 0.588  & 0.579  & -1.52 \%      & 0.221  \\
\hline 
\end{tabular}

\caption{
{\bf Selective lesion of rich-club links:} 
Summary of results for the lesion study. Selective lesion of all rich-club links leads to a decrease in functional complexity $C(lesion)$ compared to the complexity of the original network $C(real)$. After comparison to equivalent random lesions, the probability of finding a lesioned network with complexity lower than $C(lesion)$ is null for the \emph{C. elegans} and the macaque connectomes, and significantly small for the cat.
\label{tab:Lesions}
} % end caption
} % end centering
\end{table}

The remaining question is whether the observed decrease is due to the selective removal of rich-club links, or a natural consequence of perturbing the network by lesioning links. To test this we performed random lesions removing the same number of links from each SC. The rich-club links were excluded from the random lesions. We generated 100,000 realisations for each SC. In the cases of the \emph{C. elegans} and of the macaque connectomes we find that none of the randomly lesioned networks had lower complexity than the selectively lesioned SC. In the case of the cat's connectome, only $1.5\%$ of the randomly lesioned networks resulted in lower complexity. In the human SC, $22\%$ of the randomly lesioned networks lead to lower complexity than the targeted lesion. These results confirm that the rich-club is also an important feature for the functional complexity in the networks. In all cases the selective removal of rich-club links led to a measurable decrease in $C$, which resulted significant in three of the four datasets.

\begin{figure}[!ht]
\begin{center}
\includegraphics[width=3.27in]{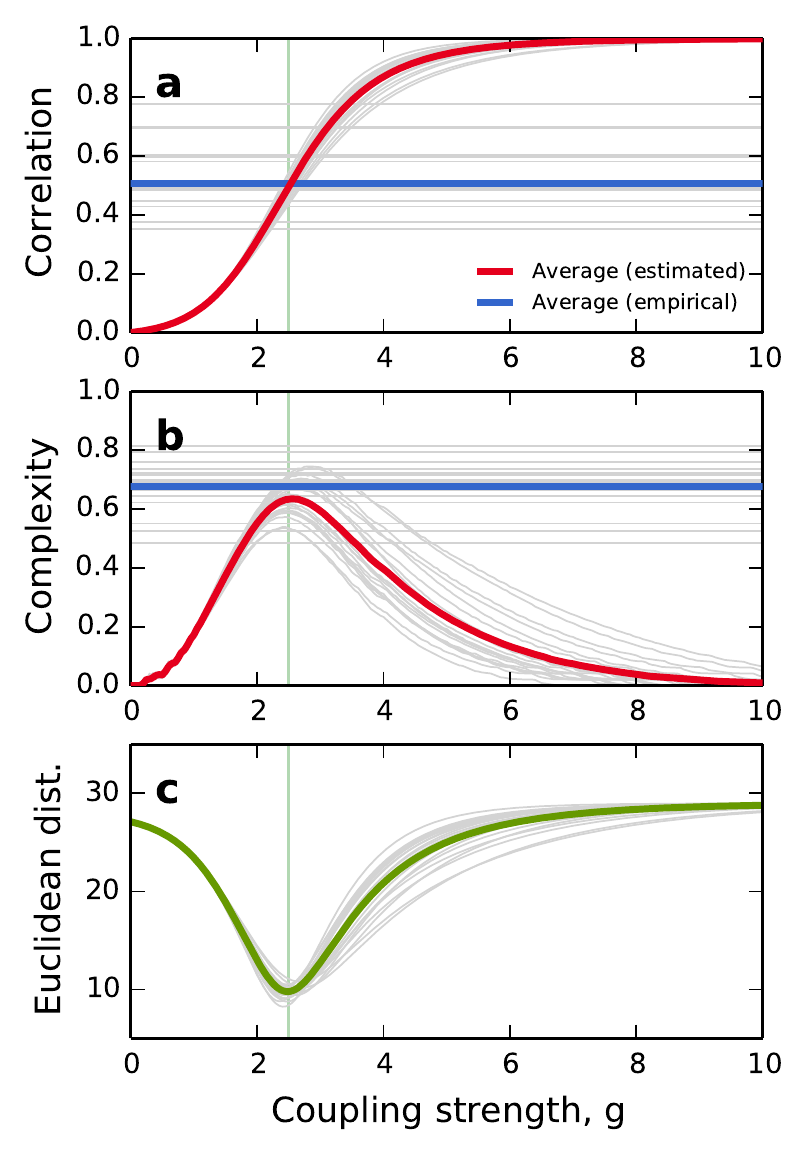}
\end{center}
\caption{
{\bf Comparison between estimated and empirical functional connectivity in human resting-state.} ({\bf a}) Mean correlation and ({\bf b}) functional complexity of simulated and empirical functional connectivity (FC) matrices. Horizontal lines are the results from empirical FC (one value per subject). Given empirical structural connectomes (tractography) corresponding FC matrices were estimated for increasing coupling $g$. Bold lines are population averages of the individual results in gray. ({\bf c}) Euclidean distance between the theoretical FCs and the empirical FCs at different values of $g$. Green bold curve is the population average. Vertical lines in the three panels mark the $g$ at which the fit is best.
}%end of caption
\label{fig:HumanRS}
\end{figure}

\begin{figure*}[!ht]
\begin{center}
\includegraphics[width=6.86in]{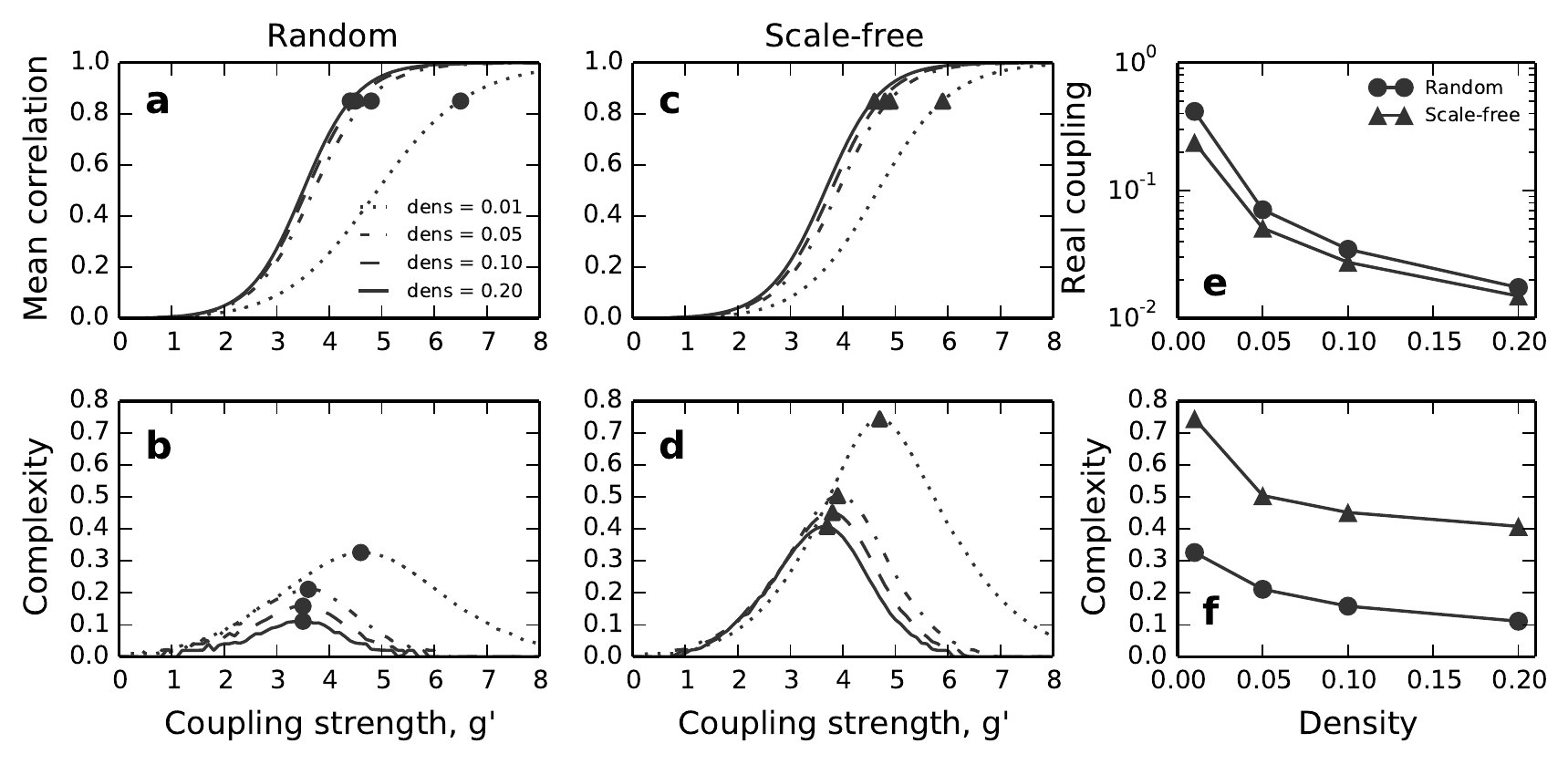}
\end{center}
\caption{
{\bf Functional complexity of random and scale-free networks.}
({\bf a}) Mean correlation and ({\bf b}) functional complexity for random graphs of $N=1000$ nodes. ({\bf c}) and ({\bf d}), same for scale-free networks. ({\bf e}) Coupling strength required for the network to reach average correlation $\left< r \right> = 0.85$ depends on the density of the network. The real coupling is $g^{real} = g \, / \, \lambda_{max}$, where $\lambda_{max}$ is the largest eigenvalue. ({\bf f}) Peak complexities reached by the networks depends on link density. All results are averages of 100 realisations.
} %end of caption
\label{fig:RandomBA}
\end{figure*}

%________________________________________________________________________________________
\subsection{Complexity of human resting-state FC}

To finish this section on the complexity of neural connectomes we turn our attention to the human connectome. We compare the functional complexity for theoretically estimated FCs and empirically obtained FCs. First, we consider SC matrices for 21 participants obtained through diffusion imaging and tractography. We estimate the theoretical FCs applying the exponential mapping to the SCs of every participant. As before, we scan for the whole range of couplings $g$. The evolution of the average correlations and the corresponding functional complexity for each participant are shown in Figs.~\ref{fig:HumanRS}(a) and (b), solid gray curves. The population averages are represented by the red solid curves. Second, we obtained empirical FC matrices for a cohort of 16 subjects via resting-state functional magnetic resonance, see Materials and Methods. Mean correlations and functional complexity were calculated out of the empirical FCs, solid horizontal lines in Figs.~\ref{fig:HumanRS}(a) and (b). The blue solid lines represent the population averages for the empirical values. Comparing the theoretical estimates and the empirical observations we find that the functional complexity of the human brain at rest lies, within the limitations of cross-subject variability, at the peak functional complexity the anatomical SCs gives rise to. Moreover, we note that this intersection happens at the coupling strength at which the simulated FCs fit closest the empirical FCs, Figure~\ref{fig:HumanRS}(c). Here we have quantified closeness as the Euclidean distance between the theoretical FC and the empirical FC matrices, diagonal entries ignored. \\

In this section we have shown that the combination of modular architecture and hubs forming rich-clubs in anatomical connectomes are key ingredients for their high functional complexity. We have also found that, within the constrains of the simple diffusive model here employed and of the cross-subject variability, the human brain at rest appears to lie in a dynamical state which matches the largest complexity that the underlying anatomical connectome can host. In the following, we want to better understand how those anatomical features give rise to larger functional complexity. Therefore, we study and compare the complexity of several benchmark graph models.

%########################################################################################
\section{Functional complexity of synthetic network models}

In this section we study the functional complexity of common synthetic network models: random, scale-free and hierarchical. We also introduce a new model of hierarchical networks which is inspired by the properties of neural and brain networks. As in the previous section, for each network we first estimate the expected correlation matrix $R$ applying the exponential mapping (see Material and Methods, Eqs.~(\ref{eq:COV}) and~(\ref{eq:Qexp})) and then we calculate the functional complexity $C$ out of the $R$ matrices using Eq.~(\ref{eq:Complexity}).

\begin{figure*}[!ht]
\begin{center}
\includegraphics[width=6.86in]{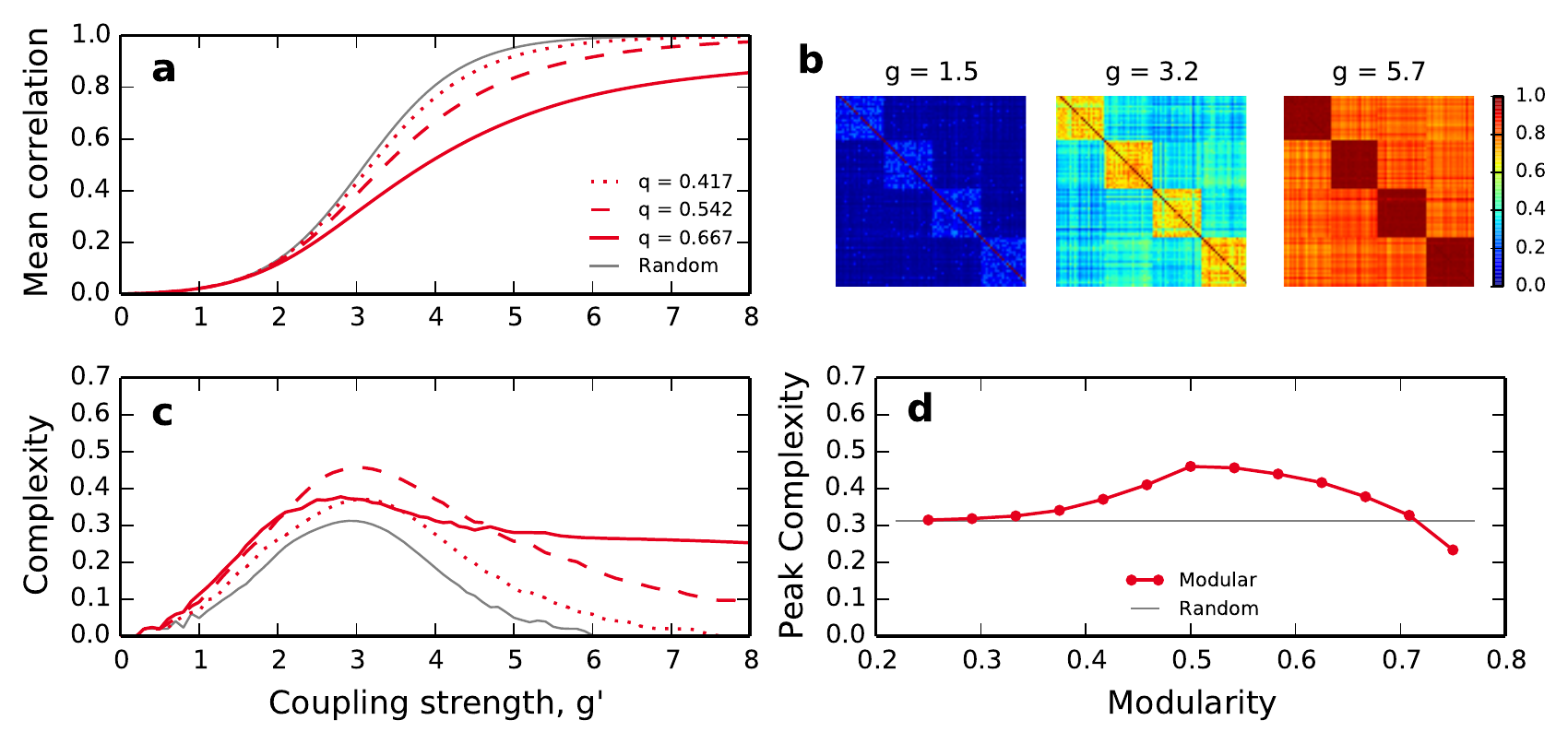}
\end{center}
\caption{
{\bf Functional complexity of modular networks.}
({\bf a}) Average correlation and ({\bf c}) functional complexity of modular networks as coupling strength $g$ increases. Networks of $N = 256$ nodes divided into 4 communities. Results for networks of different modularity (ratio of internal to external links) are shown, all with same number of links. ({\bf b}) Example correlation matrices $R$ (blue is for $r_{ij} = 0$ and red for $r_{ij} =1$). Data for one network with $q=0.542$. ({\bf d}) Peak complexity achieved by modular networks depends on modularity, compared to peak complexity of random graphs (gray line). All data points are averages of 100 realisations.
}%end of caption
\label{fig:Modular}
\end{figure*}

%________________________________________________________________________________________
\subsection{Random and scale-free networks}

We begin studying random and scale-free graphs of $N = 1000$ nodes and link densities $\rho = 0.01, \, 0.05, \, 0.1$ and $0.2$. As expected, the average correlations $\left< r \right> = \frac{2}{N(N-1)} \sum_{i =1}^N \sum_{j>i}^N r_{ij}$ increases monotonically with coupling strength $g$, Figs.~\ref{fig:RandomBA}(a) and (c), reflecting the transition the networks undergo from independence to global synchrony. Full circles ($\bullet$) and full triangles ($\blacktriangle$) mark the coupling at which $\left< r \right> = 0.85$. Considering the real coupling strength before normalisation, we see that dense networks are easier to synchronise; they reach $\left< r \right> = 0.85$ at weaker coupling, Fig.~\ref{fig:RandomBA}(e). Functional complexity always peaks in the middle of the transition, when $\left< r \right> \approx 0.5$, Figs.~\ref{fig:RandomBA}(b) and (c). The complexity of scale-free networks is notably higher than that of random graphs. The reason for this difference is that in scale-free networks the hubs synchronise with each other earlier than the rest of the nodes~\cite{Zhou_HierarchicalSynch_2006, Pereira_HubSynch_2010}. Therefore, at intermediate values of $g$ a synchronised population (composed by the hubs) coexists with the rest of nodes which are weakly correlated. See the correlation matrices in Supplementary Fig.~S2. Finally, we observe that the peak complexity decreases with density in both random and scale-free graphs, Fig.~\ref{fig:RandomBA}(f).

%________________________________________________________________________________________
\subsection{Modular networks}

We now generate networks of $N = 256$ nodes arranged into four modules of 64 nodes. Both the internal and the external links are seeded at random. We compare networks of varying modular strength by tuning the ratio of internal to external links while conserving the total mean degree to $\left< k \right> = 24$. Mean internal degree $k^{int}$ is varied from 12 to 24 and the mean external degree $k^{ext}$ is varied accordingly from 12 to 0. The strength of the modular organisation is quantified by the modularity measure $q$~\cite{Newman_Modularity_2004}. When ($k^{int}, \, k^{ext}) = (12,12)$ the network is almost a random graph. When $k^{int}$ increases (and $k^{ext}$ decreases) the modules turn stronger until they become disconnected at ($k^{int}, \, k^{ext}) = (24,0)$.

Figure~\ref{fig:Modular}(a) shows that the larger the modularity, the stronger is the coupling required to globally synchronise the network. The modules internally synchronise at rather weak couplings but to synchronise the modules with each other requires further effort, Fig.~\ref{fig:Modular}(b). The sparser the connections between the modules, the more difficult it is to synchronise them. As a consequence the distribution of correlations takes a bimodal form (see also Supplementary Fig.~S2) with one peak corresponding to the weak cross-modular interactions and a second peak for the stronger within-modular correlations. The behaviour of complexity is rather different and does not monotonically increase with modularity, Figures~\ref{fig:Modular}(c) and (d). In agreement with previous observations in modular networks of coupled phase oscillators~\cite{Zhao_Competition_2011, Adjari_Crossover_2012}, we find an optimal ratio of internal to external degrees for which complexity maximises. In our case this happens for the networks with $\left( k^{ext}, k^{int}\right) = (5,19)$ and modularity $q = 0.50$.

%________________________________________________________________________________________
\subsection{Hierarchical and modular networks}

We finish the section studying the complexity of hierarchical and modular (HM) networks. We compare three models; the first two are well known in the literature and we introduce a new model which is motivated by the properties of real brain networks. Additionally, we will compare the results to those of equivalent random and rewired networks conserving the degrees.

\begin{figure*}[!ht]
\begin{center}
\includegraphics[width=6.83in]{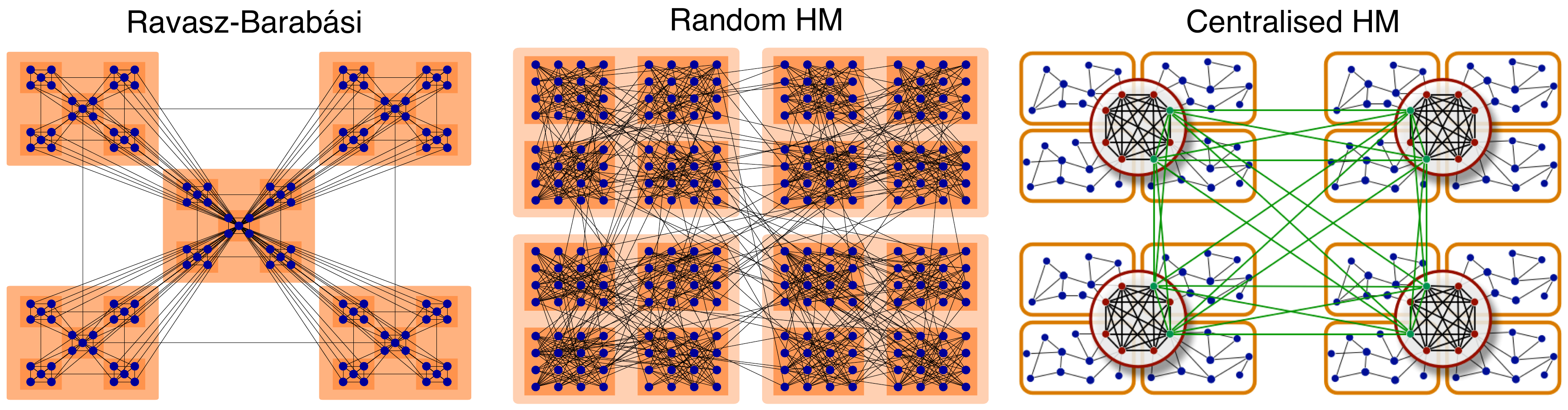}
\end{center}
\caption{
{\bf Hierarchical and modular network models:} The Ravasz-Barab\'asi model is a fractally hierarchical structure that was proposed to reproduce hierarchical features of many natural networks which posses both modules  and a scale-free degree distribution. The Random HM model by Arenas, D\'iaz-Guilera \& P\'erez-Vicente is a nested modular network in which sets of random subgraphs (the modules) are randomly linked to form larger communities. We introduce a new model, the Centralised HM model which combines a nested modular hierarchy with a scale-free-like degree distribution. This is achieved by centralising inter-modular connections through hubs that form local and global rich-clubs.
}%end of caption
\label{fig:HMnets}
\end{figure*}

%________________________________________________________________________________________
\subsubsection*{Fractally hierarchical and modular networks}

\begin{figure*}[!ht]
\begin{center}
\includegraphics[width=6.83in]{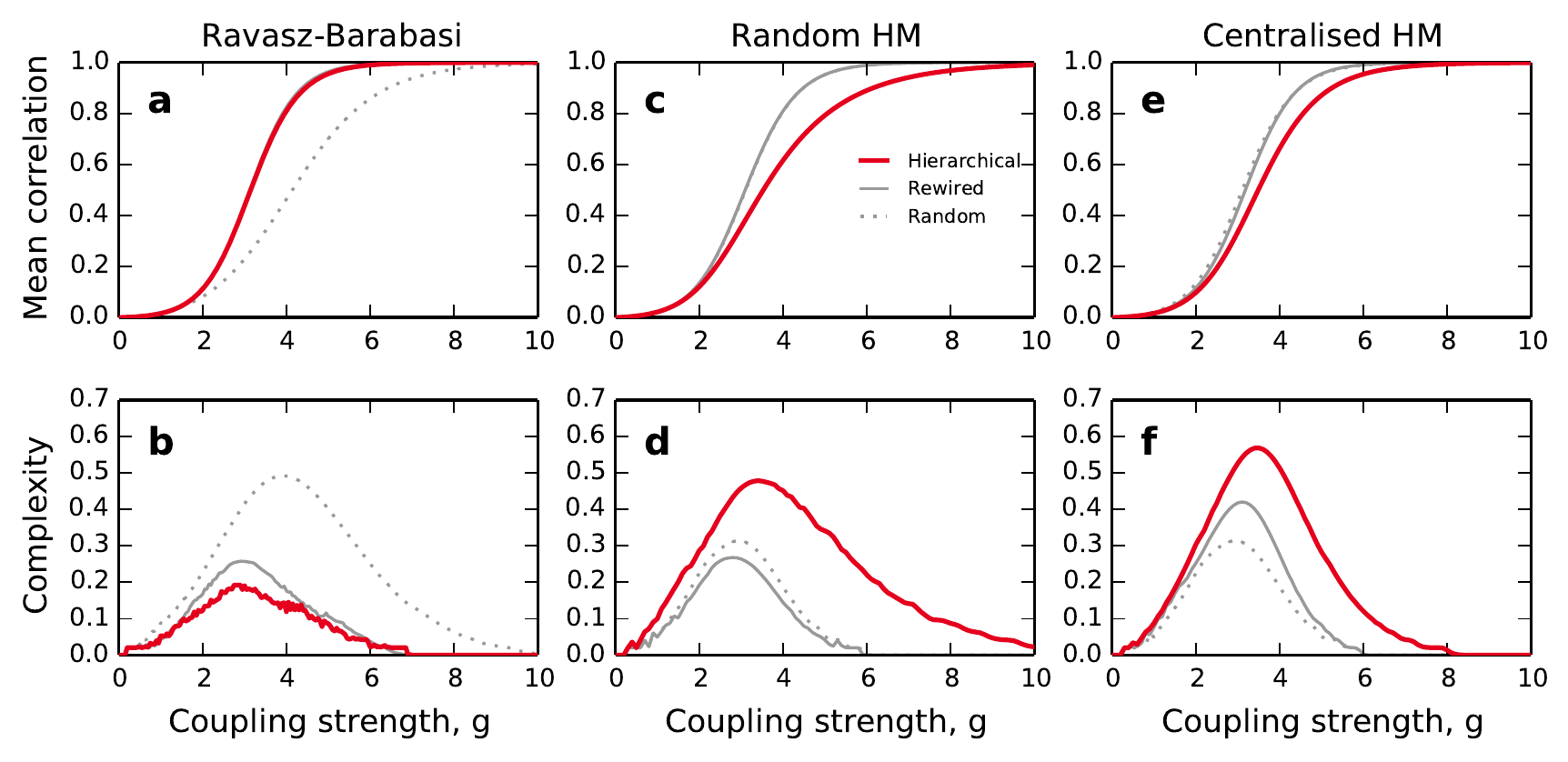}
\end{center}
\caption{
{\bf Functional complexity of hierarchical and modular (HM) networks.}
Mean correlation (upper panels) and functional complexity (lower panels) of three hierarchical and modular network models as the coupling strength $g$ is increased. (a) and (b) Ravasz \& Barab\'asi model, (c) and (d) Random HM model and, (e) and (f) Centralised HM network model. All curves are averages for 100 realisations. Variances were very small in all cases and are omitted. For each realisation of the RHM and the CHM models, 100 random and 100 rewired networks were generated.
}%end of caption
\label{fig:HM}
\end{figure*}

In an attempt to combine the modular organisation and the scale-free degree distribution found in metabolic networks, Ravasz and Barab\'asi proposed a tree-like, self-similar network model\cite{Ravasz_Metabolic_2002, Ravasz_Hierarchical_2003}. The generating motif of size $N_0$ is formed of a central hub surrounded by a ring of $N_0 -1$ nodes. To add hierarchical levels, every node is replaced by such a motif in which the original node becomes a local hub. Finally, to achieve a scale-free-like degree distribution the hubs are connected to all non-hub nodes at the lower branches. The example shown in Fig.~\ref{fig:HMnets} is the version with $N_0 = 5$ and three hierarchical levels with a total of 125 nodes. For the calculations we consider a version with $N_0 = 6$ and three hierarchical levels leading to a total size of $N=216$. Due to the deterministic nature of the model, this is the closest we can approximate to the 256 nodes of the other hierarchical networks we study. The evolution of the average correlation $\left< r \right>$ and of the complexity $C$ are shown in Figures~\ref{fig:HM}(a) and (b). The mean correlation of the Ravasz-Barab\'asi network does not distinguish from that of the rewired networks. The model achieves a very poor complexity which is overcome by both the random and the rewired networks. The large complexity of the random networks in this case can be explained by its sparse density (see Fig.~\ref{fig:RandomBA}) of only $\rho = 0.031$. The reason for why the Ravasz-Barab\'asi model  fails to match even the complexity of the rewired networks, despite having a scale-free-like degree distribution is because, by construction, the hubs of the model are preferentially connected to the non-hub nodes at the lower branches. This choice leads to a situation in which the hubs are poorly connected with each other, contrary to what happens in many real networks whose hubs form rich-clubs. The Ravasz-Barab\'asi model, on the contrary, lacks of a rich-club (see Supplementary Fig.~S7). 

%________________________________________________________________________________________
\subsubsection*{Random hierarchical and modular (RHM) networks}
In Refs.~\cite{Arenas_SynchScales_2006, Arenas_Multiresolution_2008}, Arenas, D\'iaz-Guilera and P\'erez-Vicente introduced a hierarchical network model in which a network of $N = 256$ nodes is divided into four modules of $64$ nodes, each subdivided into another four submodules of $16$ nodes, see Fig.~\ref{fig:HMnets}. The links within and across modules at all levels are shed at random. The hierarchy is defined by the increasing density at the deeper levels. In the previous section we found that in a network composed of four modules of $64$ nodes complexity was optimised when the mean external and internal degrees were $\left( k^{ext}, k^{int}\right) = (5,19)$. Taking these as the starting point we set the mean degree of the nodes at the first level to be $k^1 = k^{ext} = 5$. The remaining $19$ links are distributed among the two deeper levels. The combination $k^2 = 6$ and $k^3 = 13$ maximises the functional complexity.

The average correlation and the functional complexity of the model are shown in Figs.~\ref{fig:HM}(c) and (d). The behaviour of the Random HM networks is very similar to that of modular networks. The transition to global synchrony is governed by the interaction between the four large modules because  synchrony between the small submodules is easily achieved; see corresponding correlation matrices in Supplementary Fig. S2. The largest complexity reached by the model is $C = 0.48$, only slightly above the one of the similar modular network.

%________________________________________________________________________________________
\subsubsection*{Centralised hierarchical and modular (CHM) networks}
The coexistence of modules and scale-free-like degree distributions is a rather general observation in empirical networks. However, a model that adequately combines both features is missing. In brain connectomes both features are combined through the presence of a rich-club on top of the modular organisation. That is, cross-modular connections are not fully random but tend to be centralised through the hubs~\cite{Zamora_CatChaos_2009, Zamora_Hubs_2010, Zamora_FrontReview_2011}. Hence, we now propose a hierarchical network model which combines both features, modules and hubs, inspired by the observations in brain connectomes. For that we modify the Random HM model and replace the random connectivity between modules by a preferential attachment rule. This is achieved by sorting the nodes within a module and assigning them a probability to link with external communities proportional to their rank. See Materials and Methods for details. In the following we set the inter-modular links to be seeded with exponent $\gamma^2 = 2.0$ and the links between the four major modules (at the top level) to be placed with $\gamma^1 = 1.7$. These values for the exponents are chosen such that hubs in the resulting networks have rich-clubs comparable to those in brain connectomes, see Supplementary Fig.~S7.

The average correlation and the functional complexity of the Centralised HM model are shown in Figures~\ref{fig:HM}(e) and (f). The peak complexity is $C = 0.57$, overcoming those of the other HM networks here investigated. Also, its decay at the strong coupling regime is slower than that of random and rewired networks, as observed in the empirical brain connectomes, indicating that the model is robust against accidentally falling into global synchrony. To keep the network away from global synchrony is a desirable feature for many real systems, specially for brain connectomes. \\

So far, we have shown that a network model constructed with the topological features of empirical neural networks, a combination of modular structure with hubs centralising the cross-modular connections, achieves larger complexity than standard hierarchical network models. Specially relevant is the improvement over the fractal model by Ravasz and Barab\'asi, which has been the only network model proposed so far to combine the modular organisation and the presence of hubs in biological networks.

%########################################################################################
\section{Summary and discussion}
In the present paper we have investigated the richness of collective dynamics that networks of different characteristics can host. For that, we have proposed a measure of functional complexity based on the variability in the strength of functional interactions between the nodes. It captures the fact that complexity vanishes in the two trivial extremal cases: when the nodes are independent of each other and when the network is globally synchronised. Functional complexity emerges at intermediate states, when the collective dynamics spontaneously organise into clusters which interact with each other.
 
First, we have found that perturbation of brain's connectivity such that its modular structure is destroyed while the degrees are conserved, and the other way around, leads to networks with reduced functional complexity. The result is in agreement with the observation that rich clubs increase the set of attractors in a network of spin-glass elements beyond a scale-free topology~\cite{Senden_RichClub_2014}. We also find that the regime of high complexity is stable and robust against the network accidentally shifting towards global synchrony.

Second, we have compared the theoretically estimated functional connectivity to empirical resting-state functional connectivity in humans. We have found that, within the limitations of the nonlinear mapping here employed, the human brain at rest matches the largest functional complexity that the underlying anatomical connectome can host. This carries profound implications for understanding the relationship between structural and functional connectivity. Although the origin and the detailed role of the resting-state dynamics are still debated, it is well-known that the resting-state activity is highly structured into spatio-temporal patterns~\cite{Betzel_Repertoire_2012}. There is wide agreement that both consciousness and cognitive capacities benefit from the presence of a large pool of accessible states and to the ability to switch between them. This is supported by the finding that the dynamical repertoire of the brain is drastically decreased during sleep~\cite{Deco_CortexAsleep_2013} and under anaesthesia~\cite{Hudetz_SpinGlass_2014, Hudetz_Unconsciousness_2014}. The variability of brain signals have been found to increase with age from childhood to adulthood~\cite{McIntosh_BehaviouralVariability_2008}.

Last, but not least, we have introduced a new graph model of hierarchical and modular networks that leads to higher functional complexity than any of the models previously proposed and commonly used as benchmarks. Our model succeeds where previous efforts have failed: to combine nested modules with highly connected nodes. Specially remarkable is the improvement over the fractal model by Ravasz and Barab\'asi, which was  introduced to explain the co-existence of modular and scale-free-like degree distribution in biological networks~\cite{Ravasz_Metabolic_2002, Ravasz_Hierarchical_2003}. The model fails to foster complex dynamics; it is even outperformed by comparable random and rewired graphs. The reason is that, by construction, the hubs of the Ravasz-Barab\'asi networks are disassortative. That is, they are poorly connected with other hubs. This is contrary to the observations in brain connectomes whose hubs tend to be assortative and form rich-clubs (see Supplementary Fig.~S7). Our hierarchical network model solves the problem by centralising the cross-modular communications through hubs with a preferential attachment rule.

%________________________________________________________________________________________
\subsection*{Complexity, modules, hubs, integration and segregation}

The idea that cortical function is a combination of specialised processing by segregated neural components and their subsequent integration is an old concept in neuroscience~\cite{Damasio_BrainBinds_1989, FusterBook_2003}. For example, the Global Workspace Theory by Baars postulated the integration into a global workspace of the information processed in parallel by specialised sensory systems~\cite{Baars_Book_1988}. The lack of whole-brain structural and activity data restricted the discussions to a theoretical ground for decades. During the 1990s, the study of empirical long-range connectivity in the cat's and macaque's cortex evidenced that corticocortical connectivity is modular. Regions specialised in a sensory modality are more often interconnected than with regions of other modalities~\cite{Hilgetag_Clusters_2000, Hilgetag_ClusteredOrganiz_2004}.

The question of which is the optimal network structure that allows the brain to optimally segregate and integrate information was investigated by Tononi and Sporns in a series of network optimisation studies. Initially random networks were mutated -- slightly rewired -- and selected to maximise a cost function refered as `Neural Complexity'~\cite{Tononi_Complexity_1994, Sporns_Classes_2001}. This procedure gave rise to modular networks as those empirically observed. While modular organisation is a signature of segregation, the integration of information in modular networks can only happen via the global synchrony of the network. Global synchrony is both a rather inefficient strategy to integrate information and an undesirable state of the brain. To solve the puzzle it was proposed, again within the framework of the Global Workspace Theory, that integration might be performed by interconnected hubs which have access to the information in different segregated modules~\cite{Dehaene_NeuronalModelGW_1998, Dehaene_SubjectiveReports_2003, Shanahan_InnerLife_2010}. Closer analysis of the long-range connectivity in the cat's cortex confirmed that corticocortical connectivity is indeed organised as a modular structure with a set of rich-club hubs centralising the multisensory communications~\cite{Zamora_Thesis, Zamora_CatChaos_2009, Zamora_Hubs_2010}. Similar architectures have also been identified in the human anatomical connectome~\cite{Heuvel_HubsHuman_2011} and in the neural architecture of the \emph{Caenorhabditis elegans}~\cite{Varshney_Elegans_2011}. These findings are the starting point for the hypothesis that the specialisation of cortical regions to different sensory modality may have triggered the segregation of cortical regions into network modules, while the rich-club hubs may be the responsible for the integration step~\cite{Zamora_Thesis, Zamora_Hubs_2010, Zamora_FrontReview_2011, Shanahan_InnerLife_2010, Sporns_AttributesIS_2013}. We shall notice that this scenario, although reasonable and plausible, is still a \emph{working hypothesis} which needs empirical demonstration. Results on the functional organisation of the brain during rest and task start to support the hypothesis~\cite{Bartolero_IntegArchitecture_2015}. The identification alone of the modular architecture with rich-clubs does neither explain \emph{how} does the brain perform the integration step. Because integration is inherently related to the dynamic nature of the sensory inputs~\cite{Deco_Rethinking_2015}, models of propagation of information in networks will become relevant to understand this question in the future~\cite{Misic_CooperativeSpreading_2015}. 

A common source of confusion in the literature arises from the implicit causal relationships which have been drawn between the network's structural features, its complexity and its capacity to integrate and segregate information. The network optimisation performed by Tononi \& Sporns illustrates some of these limitations. On the one hand, the procedure led to modular networks without hubs. Thus it remains an open question whether optimisation of complexity alone results in the network topologies we observe empirically. It would be of high interest to use multidimensional optimisation methods which can account for several constraints~\cite{Avena_ParetoOptimality_2014} to clarify this question. Even if the results were negative, that optimisation of complexity alone does not generate hierarchical networks with modules and hubs, this would be important to better understand the ``driving-forces'' which shaped the brain's connectivity along evolution.
On the other hand, Tononi and Sporns presupposed that maximisation of the \emph{neural complexity} measure implied an increase of the network's capacity to integrate and segregate information. However, this causal relation has never been demonstrated. When comparing to our Functional Complexity measure, we have found that neural complexity monotonically increases with coupling strength and is maximal when the networks are globally synchronised, see Supplementary Information. This contradicts the intention of the measure, which was aimed at identifying networks with optimal balance between integration and segregation. At the globally synchronised state there is no segregation and hence, no optimal balance. In order to overcome this limitation Zhao, Zhou and Chen proposed an alternative measure of complexity based on the entropy of the distribution of cross-correlation values~\cite{Zhao_Complexity_2010, Zhao_Competition_2011}. Here, we have adopted their approach but we have replaced the nonlinear entropy function by the integral between the observed and the uniform distributions. This choice considerably enhances the discriminative power of the measure and its robustness to variation in the number of bins to estimate the distribution. Find a comparison in the Supplementary Information.

Summarising, in this paper we could confirm that the hierarchical organisation of networks into modules interconnected via rich-club hubs lead to an increase of their functional complexity. As we have argued, the opposite might not be true. This, and the precise causal relations between a network's complexity and its capacity to integrate and segregate information are still open questions which demand clarification. Explicit efforts are required within the field of brain connectivity to corroborate or discard the validity of these \emph{working hypothesis}. In the mean-time, we shall refrain from taking these causal relations for granted; specially, when applying such concepts to evaluate and interpret clinical conditions~\cite{Koch_NeuralCorrelates_2016, Tononi_Consciousness_2016}.

%________________________________________________________________________________________
\subsection*{Limitations}

The temporal evolution of the functional complexity is a relevant aspect we have ignored here. A complete characterisation still requires further developments because the complexity of a coupled dynamical system is composed by two aspects. One is the formation of complex coalitions between the nodes. This is the  \emph{interaction complexity} that we have studied here and which we have coined as \emph{functional complexity}. The other aspect is the complexity of the time-courses traced by the individual signals, or by the system as a whole. From a temporal perspective, neither random nor periodic signals are complex~\cite{Wackerbauer_Complexity_1994}. Random signals are unpredictable but represent fully disordered behaviour. Periodic signals are predictable but represent ordered behaviour. On the other hand, chaotic signals are complex because they are the result of an intricate mixture of order and unpredictability. However, we should notice that a set of chaotic elements may be synchronised giving rise to low functional complexity; a set of coupled periodic signals could result into heterogeneous spatial correlations due to the network's topology leading to high functional complexity.

The results here presented are to be considered under the constraints and the limitations of the heuristic exponential mapping we have introduced to theoretically estimate the networks' functional connectivity. ($i$) The model is an estimation of the time-averaged cross-correlation and is therefore not suitable to evaluate temporal fluctuations. ($ii$) In diffusion models the nodes are considered as passive relay stations for the flow of information while brain regions and neurones likely perform nonlinear transformations to the incoming information through active and complex dynamics at the local circuits.

Neural dynamics emerge from nonlinear interaction together with external stochastic perturbations, characterised by complex oscillations. Detailed collective dynamics may depend on various factors, including nonlinearity in the local dynamics and the form of the interactions. However, some generic properties may not be dominantly determined by such details, especially the dependence of the dynamical correlation on the network topology. Dynamical cross-correlation (functional connectivity) measures the interdependence between brain areas over relatively long time scales, thus are mainly determined by the slow modes which can be captured by the first few modes in the expansion of the local dynamics and coupling functions~\cite{Galan_DominantPatterns_2008}. Estimations derived from linear Gaussian processes make good sense to capture the correlated fluctuations around the leading linear modes, but it diverges at strong couplings because it ignores the higher order dynamics. To gain some understanding, we have illustrated how a Gaussian diffusion process can be interpreted in terms of the networks' graph properties, in particular on the role played by communication paths of different length.

The heuristic exponential mapping we have proposed further improves such a strategy by effectively taking some high-order contributions into consideration; specially, the decay of information along the path. It accounts for the fact that information in a neural system is used or transformed instead of perpetually diffuse within the network. This avoids the divergence problems of the linearised Gaussian diffusion process without the need to consider a particular dynamical model. This is also the reason for why the simple exponential mapping can capture the behaviour of some coupled dynamical models, e.g., Kuramoto and neural mass models (see Supplementary Figure S6). An advantage of the exponential mapping for the exploratory purposes of the current paper was the computational efficiency. It would have been impractical to perform the extensive set of comparisons with surrogate networks if actual simulations were to be run for every datapoint.

%________________________________________________________________________________________
\subsection*{Outlook}
From a practical point of view our measure of functional complexity is an excellent candidate as a clinical marker for connectivity-related conditions. Over the last decade it has been consistently reported how resting-state functional connectivity differs across healthy subjects and patients suffering from diverse conditions~\cite{Fox_RSapplications_2010}. Most of those reports are based in the graph analysis of functional connectivity which unfortunately depend on several arbitrary choices~\cite{Fornito_Review_2013, Papo_GreatExpectations_2014}, e.g., the need to set a threshold to binarise the correlation matrix. Our measure of complexity requires no unreasoned choices, it is easy to apply and interpret. The measure can be applied to any metric of pair-wise functional connectivity, e.g., mutual information, despite we restricted here to cross-correlation. It is also very fast to compute and is thus suitable for real-time monitoring systems.

The new hierarchical graph model we have introduced represents a very satisfactory compromise to combine hierarchically modular architecture with a broad degree distribution. From a biological point of view, however, we recognise the need to define models which can explain how brain networks developed their current topology in the course of evolution. We foresee that the key ingredients of such evolutionary models are: ($i$) identification of the ``driving-forces'', e.g., the balancing between integration and segregation, and ($ii$) a growth process that accounts for the increase in the number of neurones or cortical surface over time~\cite{Kaiser_Development_2004, Kaiser_SpatialGrowth_2004}. Other ingredients shall include ($iii$) the trade-off between the cost and the efficiency of the resulting networks given the spatial constraints of the brain~\cite{Kaiser_Placement_2006, Chen_TradeOff_2013, Samu_WiringCost_2014} and ($iv$) the patterns of axonal growth during development~\cite{Lim_AxonGrowth_2015}. 

As a final remark, we shall notice that although we have here restricted to the study of neural and brain connectomes, we are confident that the modular organisation with centralised intercommunication is a general principle of organisation in biological networks. We find it reasonable that the assumption of balancing between integration and segregation as the principal driving-force to shape the large-scale neural connectomes, is also applicable to other networked biological systems. For example, we recently reported that the transcriptional regulatory network of the \emph{Mycobacterium tuberculosis} shares fundamental properties with those of neural neural networks~\cite{Klimm_Roles_2014}. In the end, the transcriptional regulatory network is the system responsible in the small bacterium to collect information of the environment through different channels, to process and interpret that information, and to efficiently combine it to ``take decisions'' that improve the chances of survival.

%########################################################################################
\section{Materials and Methods}

\small
%________________________________________________________________________________________
\subsection{Connectivity datasets}

\begin{table*}
\centering
\begin{tabular}{l c c c c | c c}
Network 			& $N$ & $L$ & Density & Reciprocity & Comms. & Modularity  \\
\hline
\emph{C. elegans}	& 275 & 2990 & 0.04 & 0.47 & 4 &  0.417 \\
Cat 				& 53 & 826 & 0.30 & 0.73 & 4 & 0.270 \\
Macaque			& 85 & 2356 & 0.33 & 0.74 & 3 & 0.402 \\
Human (21 subs)	& 76 &  655 - 1061 & 0.23 -0.37 & undir. & -- & ---- \\
Human (average)	& 76 & 935 & 0.33 & undir. & 3 & 0.33 \\
\hline
\end{tabular}

\caption{{\bf Summary of neural and brain network datasets:} In this paper we have used four real datasets of structural connectivity. The table sketches their principal properties: their size $N$ (number of neurones or cortical areas), their number of links $L$ or the number of communities identified (Comms.). The human structural connectivity is the only undirected dataset because tractography does not distinguish directionality of the projections.
}
\label{tab:NetSummary}
\end{table*}

\noindent {\bf Caenorhabditis elegans:}
The \emph{C. elegans} is a small nematode of approximately 1~mm long and is one of the most studied organisms. Its nervous system consists of 302 neurones which communicate through gap junctions and chemical synapses. We use the collation performed by Varshney et al. in Ref.~\cite{Varshney_Elegans_2011}; the data can be obtained in http://wormatlas.org/neuronalwiring.html. After organising and cleaning the data we ended with a network of $N = 275$ neurones and $L = 2990$ links between them. For the general purposes of the paper we consider two neurones connected if there is at least one gap junction or one chemical synapse between them. We ignored all neurones that receive no inputs because they are always dynamically independent. The resulting network has a density of $\rho = 0.04$ and a reciprocity of $0.470$ meaning that $47\%$ of links join neurones A and B in both directions while the remaining $53\%$ connect two neurones in only one direction. Most of the reciprocal connections come from the gap junctions, which are always bidirectional; only $21\%$ of the chemical synapses are devoted to connect two neurones in both directions.

\noindent {\bf Cat cortex:}
The dataset of the corticocortical connections in cats was created after an extensive collation of literature reporting anatomical tract-tracing experiments~\cite{Scannell1993, Scannell1995}. It consists of a parcellation into $N = 53$ cortical areas of one cerebral hemisphere and $L = 826$ fibres of axons between them.
After application of data mining methods~\cite{Scannell1993, Hilgetag_ClusteredOrganiz_2004} the network was found to be organised into four distinguishable clusters which closely follow functional subdivisions: visual, auditory, somatosensory-motor and frontolimbic. The network has a density of $\rho = 0.30$ and $73\%$ of the connections are reciprocal.

\noindent {\bf Macaque monkey:}
The macaque network is based on a parcellation of one cortical hemisphere into $N = 95$ areas and $L = 2390$ directed projections between them~\cite{Kaiser_Placement_2006}. The dataset, which can be downloaded from http://www.biological-networks.org, is a collation of tract-tracing experiments gathered in the CoCoMac database (http://cocomac.org)~\cite{Kotter_Retrieval_2004}. Ignoring all cortical areas that receive no input we ended with a reduced network of $N = 85$ cortical areas. The network has a density of $\rho = 0.33$ and reciprocity $r = 0.74$.

\noindent {\bf Human structural connectivity:}
Structural connectivity was acquired from 21 healthy right-handed volunteers. Find full details in Ref.~\cite{Cabral_Disconnection_2012, Hartevelt_DBS_2014}. This study was approved by the National Research Ethics Service (NRES) committee South Central -- Berkshire in Bristol and carried out in accordance with the approved guidelines. All healthy participants gave written informed consent.

Diffusion imaging data were acquired on a Philips Achieva 1.5 Tesla Magnet in Oxford from all participants using a single-shot echo planar sequence with coverage of the whole brain. The scanning parameters were echo time (TE) = 65 ms, repetition time (TR) = 9390 ms, reconstructed matrix $176 \times 176$ and reconstructed voxel size of $1.8 \times 1.8$ mm and slice thickness of 2 mm. Furthermore, DTI data were acquired with 33 optimal nonlinear diffusion gradient directions ($b$ = 1200 s/mm$^2$) and 1 non-diffusion weighted volume ($b = 0$). We used the Automated Anatomical Labelling (AAL) template to parcellate the entire brain into 90 cortical and subcortical (45 each hemisphere) as well as 26 cerebellar regions (cerebellum and vermis)~\cite{Tzourio_AALparcellation_2002}. The parcellation was conducted in the diffusion MRI native space. We estimated the connectivity probability by applying probabilistic tractography at the voxel level using a sampling of 5000 streamline fibres per voxel. The connectivity probability between region $i$ to $j$ was defined by the proportion of fibres passing through voxels in $i$ that reach voxels in $j$~\cite{Behrens_ProbDTI_2007}. Because of the dependence of tractography on the seeding location, the probability from $i$ to $j$ is not necessarily equivalent to that from $j$ to $i$. However, these two probabilities were highly correlated, we therefore defined the undirected connectivity probability $P_{ij}$ between regions $i$ and $j$ by averaging these two probabilities. We implemented the calculation of regional connectivity probability using in-house Perl scripts. Regional connectivity was normalised using the regions' volume expressed in number of voxels.

The 21 networks so constructed were all composed of $N = 116$ brain regions and a number of links ranging from $L = 1110$ undirected links for the sparsest case (density $\rho = 0.17$) to $L = 1614$ for the densest ($\rho = 0.24$). These networks are individually used for the results in Fig.~\ref{fig:HumanRS}. In order to derive an average connectome, used in the results of Fig.~\ref{fig:RealNets}, we performed an iterative procedure which automatically prunes outlier links (data-points falling out of 1.5 times the inter-quartile range). For each link between regions $i$ and $j$ outlier values of $C_{ij}$ are identified among the initial 21 measures available for the link (one per subject) as those values out of the $1.5$ inter-quartile range (IQR). If outliers are identified we remove them from the dataset and search again for outliers. The procedure stops when no further outliers are identified. This method allows to clean the data without having to set an arbitrary threshold for the minimally accepted prevalence of the link across subjects. The  average network contains approximately the same number of links as the individual matrices. Defining the average connectivity by computing the simple mean across the 21 $C_{ij}$ matrices (a usual approach in the literature) leads to an average connectivity matrix that contains more than twice the links in the matrices for individual subjects. For consistency with the datasets of the cats and the macaque monkeys, we show in the paper the results for the subnetworks formed only by the $N = 76$ cortical regions ($38$ per hemisphere) and ignoring all subcortical areas. We found qualitatively the same results in the cortical subnetwork and in the full-brain network, with the only difference that the cerebellum and the vermis form a very densely interconnected community that synchronises easily.

\noindent {\bf Human functional connectivity:}
Data were collected at CFIN, Aarhus University, Denmark, from 16 healthy right-handed participants (11 men and 5 women, mean age: 24.75 $\pm$ 2.54). All participants were recruited through the online recruitment system at Aarhus University excluding anyone with psychiatric or neurological disorders (or a history thereof). The study was approved by the internal research board at CFIN, Aarhus University, Denmark. Ethics approval was granted by the Research Ethics Committee of the Central Denmark Region (De Videnskabsetiske Komit\'er for Region Midtjylland). Written informed consent was obtained from all participants prior to participation.

The MRI data (structural MRI and rs-fMRI) were collected in one session on a 3T Siemens Skyra scanner at CFIN, Aarhus University, Denmark. The parameters for the structural MRI T1 scan were as follows: voxel size of 1 mm$^3$; reconstructed matrix size $256 \times 256$; echo time (TE) of 3.8 ms and repetition time (TR) of 2300 ms. The resting-state fMRI data were collected using whole-brain echo planar images (EPI) with TR = 3030 ms, TE = 27 ms, flip angle = 90$^\circ$, reconstructed matrix size = $96 \times 96$, voxel size $2 \times 2$ mm with slice thickness of 2.6 mm and a bandwidth of 1795 Hz/Px. Approximately seven minutes of resting state data were collected per subject.

We used the automated anatomical labelling (AAL) template to parcellate the entire brain into 116 regions~\cite{Tzourio_AALparcellation_2002}. The linear registration tool from the FSL toolbox (\emph{www.fmrib.ox.ac.uk/fsl}, FMRIB, Oxford)~\cite{Jenkinson_MotionCorrection_2002} was used to co-register the EPI image to the T1-weighted structural image. The T1-weighted image was co-registered to the T1 template of ICBM152 in MNI space~\cite{Collins_Automatic3D_1994}. The resulting transformations were concatenated and inverted and further applied to warp the AAL template~\cite{Tzourio_AALparcellation_2002} from MNI space to the EPI native space, where interpolation using nearest-neighbour method ensured that the discrete labelling values were preserved. Thus the brain parcellations were conducted in each individual's native space.

Data preprocessing of the functional fMRI data was carried out using MELODIC (Multivariate Exploratory Linear Decomposition into Independent Components) Version 3.14~\cite{Beckmann_ProbICA_2004}, part of FSL (FMRIB's Software Library, www.fmrib.ox.ac.uk/fsl). We used the default parameters of this imaging pre-processing pipeline on all participants: motion correction using MCFLIRT~\cite{Jenkinson_MotionCorrection_2002}; non-brain removal using BET~\cite{Smith_FastExtraction_2002}; spatial smoothing using a Gaussian kernel of FWHM 5mm; grand-mean intensity normalisation of the entire 4D dataset by a single multiplicative factor and high pass temporal filtering (Gaussian-weighted least-squares straight line fitting, with sigma = 50.0s). We used tools from FSL to extract and average the time courses from all voxels within each AAL cluster. We then used Matlab (The MathWorks Inc.) to compute the pairwise Pearson correlation between all 116 regions, applying Fisher's transform to the r-values to get the z-values for the final 116x116 FC-fMRI matrix.

%________________________________________________________________________________________
\subsection{Community detection}
After application of data mining methods the corticocortical network of the cat network was found to be organised into $4$ distinguishable clusters which closely follow functional subdivisions: visual, auditory, somatosensory-motor and frontolimbic~\cite{Scannell1993, Hilgetag_ClusteredOrganiz_2004}. For the other three real datasets we investigated their modular structure using \texttt{Radatools} (http://deim.urv.cat/$\sim$sergio.gomez/radatools.php), a software that allows to detect graph communities by alternating different methods. We run the community detection such that it would first perform a coarse grained identification of the communities using Newman and Girvan's method~\cite{Newman_Modularity_2004} and then a method by G\'omez and Arenas named `Tabu Search'~\cite{Arenas_Multiresolution_2008} was applied. Final optimisation of the partitions was performed by a reposition method.

The neural network of the \emph{C. elegans} was partitioned into four modules of size 8, 64, 92 and 111 neurones respectively with modularity $q = 0.417$. The corticocortical network of the macaque monkey was divided into three modules of 4, 38 and 47 cortical areas with $q = 0.402$. The average human corticocortical connectome was divided into three modules of sizes 20, 26 and 30 with modularity $q = 0.33$. In two of the modules there is a dominance of one cortical hemisphere while the third module contains left and right areas in similar numbers.\\

%________________________________________________________________________________________
\subsection{Surrogates and synthetic network models}
The network analysis, the generation of network models and the randomisation of networks has been performed using \texttt{GAlib}, a library for graph analysis in Python (https://github.com/gorkazl/pyGAlib). The network generation and rewiring functions are located in the submodule \emph{gamodels.py}.

\noindent {\bf Random graphs:} Random graphs were generated following the $G(N,L)$ model that seeds links between randomly chosen pairs of nodes in an initially empty graph of $N$ nodes until $L$ links have been placed. Random graphs were produced using the function \emph{RandomGraph}.

\noindent {\bf Scale-free networks:} Random graphs with scale-free degree distribution were generated following the method in Ref.~\cite{Goh_LoadDistribution_2001}. The nodes are ranked as $i = 1, 2, \ldots, N$ and they are assigned a weight $p(i) = \frac{i^{-\alpha}}{\sum_j j^{-\alpha}}$. To place the links, two nodes $i$ and $j$ are chosen at random with probabilities $p(i)$ and $p(j)$ respectively and they become connected if they were not already linked. The procedure is repeated until the desired number of links are reached. Scale-free networks generated using this preferential attachment rule achieve, on the limit of large and sparse networks, a degree distribution $p(k) \propto e^{- \gamma}$ with $\gamma = (1 + \alpha) / \alpha > 2$. Tuning $\alpha$ in the range $[0,1)$ scale-free networks with exponent $\gamma \in [2, \infty)$ are generated. 
Here, we set $\alpha = 0.5$ to achieve scale-free networks with $\gamma = 3.0$. The exponent in individual network realisations fluctuated between $2.6$ and $3.4$. The function \emph{ScaleFreeGraph} generates scale-free networks with desired size $N$, number of links $L$ and exponent $\gamma$.

\noindent {\bf Rewired networks:} Given a real network it is often desirable to compare it with equivalent random graphs which have the same degree distribution as the original network. A common procedure is to iteratively choose two links at random, e.g. ($i$,$j$) and ($i'$,$j'$), and to switch them such that ($i$,$j'$) and ($i'$,$j$) are the new links. The method is usually attributed to Maslov \& Sneppen~\cite{Maslov_ProteinNets_2002} but it had been proposed by several authors before~\cite{Katz_Rewiring, Holland_1977, Rao_1996, Kannan_1999, Roberts_2000}. The function \emph{RewireNetwork} returns rewired versions of a given input network. In order to guarantee the convergence of the algorithm into the subspace of maximally random graphs, the link-switching step is repeated for $10 \times L$ iterations, where $L$ is the number of links.

\noindent {\bf Modular and nested-hierarchical networks:}
Random modular networks were generated, as the random graphs, choosing two nodes at random and connecting them if they were not previously linked. The difference lies on choosing the two nodes either from the same module or from two different modules. The nested hierarchical model with random connectivity~\cite{Arenas_SynchScales_2006} is an extension of this procedure such that modules are subdivided into further modules. The function \emph{HMRandomGraph} generates both modular and nested hierarchical networks depending on the input parameters.

\noindent {\bf Nested-hierarchical networks with centralised connectivity:}
While in the nested hierarchical random model the connections between the modules are shed at random, in neural and brain networks inter-module connections and communication paths tend to be centralised through the hubs~\cite{Zamora_CatChaos_2009}. We here propose a model of hierarchical and modular networks that combines both features. For that we modify the nested hierarchical model to replace the random connectivity between modules by a preferential attachment rule. We start by creating the $16$ random graphs of $16$ nodes each and mean degree $\kappa^3 = 13$ of the deepest level. Then, the nodes of each submodule are ranked as $i = 1, 2, \ldots, N_3 = 16$ and they are assigned a weight $p(i) = \frac{i^{-\alpha}}{\sum_j j^{-\alpha}}$. To place the inter-modular links, two nodes $i$ and $j$ are chosen at random from two different modules with probability $p(i)$ and $p(j)$ respectively, and they become connected if they were not already linked. The procedure is repeated at each hierarchical level until the mean degree of inter-modular links are $\kappa^2 = 6$ and $\kappa^1 = 5$ as we had in the nested random hierarchical networks. Scale-free networks generated using this preferential attachment rule achieve, on the limit of large and sparse networks, a degree distribution $p(k) \propto e^{- \gamma}$ with $\gamma = (1 + \alpha) / \alpha > 2$~\cite{Goh_LoadDistribution_2001}. Tuning $\alpha$ in the range $[0,1)$ scale-free networks with exponent $\gamma \in [2, \infty)$ are generated. The inter-modular links at the second level are planted using $\gamma^2 = 2.0$ and the links between the four major modules at first level are placed with $\gamma^1 = 1.7$. The function \emph{HMCentralisedGraph} generates nested-modular hierarchical networks with the inter-modular links centralised, seeding the inter-modular links at each level with a preferential attachment rule of desired exponent $\gamma$.

\noindent {\bf Modularity preserving random graphs:}
Given a network with a partition of its $N$ nodes into $n$ communities we generated graphs with the same modularity but randomly connected. Therefore we first counted in the original network the number of links $L_{rs}$ between any two communities $r, s = 1, 2, \ldots, n$. So, $L_{rr}$ are the number of internal links within the community $r$ and $L_{rs}$ are the number of links between nodes in community $r$ and community $s$. Then, the generation procedure is the same as for the random modular networks but considering the specific number of links to be planted in each case. The resulting random networks have the same modularity $q$ as the original network for the given partition. Modularity preserving random graphs were generated using the function \emph{ModularityPreservingGraph}.

\noindent {\bf Ravasz-Barab\'asi networks:}
The Ravasz-Barab\'asi model is composed of a ring of $N_0 -1$ nodes connected to their first neighbours surrounding a central hub. To generate subsequent hierarchical levels every node becomes the central node of a copy of the original motif. Finally, the central nodes are connected to all the non-hub nodes in the lower hierarchical levels of the branch they belong to. The function \emph{RavaszBarabasiGraph} creates Ravasz-Barab\'asi networks with desired size $N_0$ in the original motif and desired number of hierarchical levels.

%________________________________________________________________________________________
\subsection{Mapping functional connectivity from anatomical connectomes}
The collective dynamics of coupled systems depend on many factors such as the topology of the network (the structural connectome), the model chosen for the local node dynamics (e.g., Kuramoto oscillators, neural-mass models or spiking neurones) and the coupling function between them which determines how information is passed from one node to its neighbours. The purpose of the present paper is to investigate the influence of the connection topology while discarding as much as possible other influences. For that we propose a simple mapping to analytically estimate the functional connectivity out of a structural connectome. This is accomplished by considering a diffusive model with non-linear (exponential) decay of information transmission at longer paths. The convenience of this mapping for the present purposes are twofold. First, it avoids internal parameters. The only free parameter controlling the collective behaviour is the strength of the connections. Second, it allows to analytically estimate the functional connectivity without the need to run otherwise time-consuming simulations. 

In order to illustrate the nonlinear mapping we here propose, let us first revisit a widely used linear stochastic model. Given a network of $N$ neural populations represented by the binary adjacency matrix $A$ with $A_{ij} = 1$ if node $i$ sends a projection to node $j$ and $A_{ij} = 0$ otherwise, the firing rate $r_i$ of each population can be expressed following a generic rate equations, often also referred as the Wilson-Cowan model:
%%%%%%%%
\begin{equation}
\dot{r}_i = - \alpha \, r_i + \Theta \left(g \, \sum_{j=1}^N A_{ji} \, r_j + I_i + \eta_i \right),
\label{eq:WilsonCowan}
\end{equation}
%%%%%%%
where $\alpha$ is the inverse of the relaxation time, $\Theta(\cdot)$ is a positive sigmoidal function, $g$ the coupling strength, $I_i$ an external input and $\eta_i$ a noise term. Under the assumption of weak coupling the fluctuations $x_i$ of the firing rates $r_i$ around their mean can be linearised as:
%%%%%%%%
\begin{equation}
\dot{x}_i = - \alpha \, x_i + g \, \sum_{j=1}^N A_{ji} \, x_j + \sqrt{2 \, \alpha \, \sigma^2} \, \xi_i,
\label{eq:OUprocess}
\end{equation}
%%%%%%%
where $\xi_i$ is now a Gaussian white noise with zero mean and unit variance, and $\sigma$ its variance. This is also known as an Ornstein-Uhlenbeck stochastic process. Given the column vectors ${\bf x}^T \! = (x_1, x_2, \ldots, x_N)$ and $\boldsymbol{\xi}^T \! = (\xi_1, \xi_2, \ldots, \xi_N)$, the system can be rewritten in matrix form as:
%%%%%%%%
\begin{equation}
\mathbf{\dot{x}} = - \alpha \, \mathbf{x} + g \, \mathbf{A}^T \, \mathbf{x} + \sqrt{2 \, \alpha \, \sigma^2} \; \boldsymbol{\xi}.
\label{eq:OUmatrix}
\end{equation}
%%%%%%%
The transpose of the adjacency matrix $\mathbf{A}^T$ is important when the network is directed such that the dynamics of population $i$ is determined by its inputs, not by its outputs. The covariance of this multivariate Gaussian system can be analytically estimated~\cite{Papoulis1991, Tononi_Complexity_1994} by averaging over the states produced by an ensemble of noise vectors $\boldsymbol{\xi}$. Defining $Q = (\mathbf{1} - \, ^g\! / _\alpha A^T)^{-1}$, the covariance matrix is thus 
%%%%%%%%
\begin{equation}
COV(X) = \left< {\mathbf x} \cdot {\mathbf x}^T \right> \; = \; 
\frac{2 \sigma^2}{\alpha} \, \left< \left(Q \, \boldsymbol{\xi} \right) \cdot ( \boldsymbol{\xi}^T \, Q^T ) \right> \; = \; 
%\frac{2 \sigma^2}{\alpha} \, Q \left< \boldsymbol{\xi} \cdot \boldsymbol{\xi}^T \right> Q^T \; = \; 
\frac{2 \sigma^2}{\alpha} \, Q \cdot Q^T.
\label{eq:COV}
\end{equation}
%%%%%%%%
As stated above, the Gaussian diffusion process in Eq.~(\ref{eq:OUprocess}) is a linear approximation of the fluctuations in Eq.~(\ref{eq:WilsonCowan}) that is valid only for weakly coupled networks. Because the linear equation lacks of the sigmoidal function $\Theta(\cdot)$ which delimits the amplitude of the inputs received by a neural population, its solution diverges to infinity when $g / \alpha$ equals any of the eigenvalues of the adjacency matrix $A$~\cite{Zamora_Hubs_2010}. In order to compare networks of different size and density, the coupling strength shall be normalised such that $\tilde{g} = g \, / \, \lambda_{max}$ where $\lambda_{max}$ is the largest eigenvalue of $A$. For directed networks $\lambda_{max}$ is replaced by the largest norm of $A$'s complex eigenvalues. In this case the matrix $Q$ is defined for all $\tilde{g} \in [0, \; \alpha)$ and the solutions of Equations~(\ref{eq:OUprocess}) and (\ref{eq:OUmatrix}) converge. The $\alpha$ parameter, which is usually regarded as an important parameter that controls re-entrant \emph{self-activations} of the neural population plays here a rather irrelevant role. It only re-scales the coupling, shifting the strength at which the network diverges but it does not change the functional form of the solutions. Hence, we will consider $\alpha = 1$. We will also consider that $g$ is the normalised coupling such that the system converges for $g \in [0,\,1)$.

We now explain the solution of the Gaussian diffusion model in terms of the graph properties of the underlying structural connectivity. Therefore we note that the matrix Q can be represented as the following series expansion:
%%%%%%%%
\begin{equation}
Q = \frac{1}{\mathbf{1} - gA} = \mathbf{1} + g A + g^2 A^2 + g^3 A^3 + \cdots = \sum_{l=0}^\infty g^l A^l.
\label{eq:Qmatrix}
\end{equation}
%%%%%%%%
It is well-known that in a network the total number of paths of length $l$ between two nodes is given by the powers of the adjacency matrix $A^l$. This includes paths with internal recurrent loops. From this point of view we realise that $Q_{ji}$ represents the total \emph{influence} exerted by node $j$ over node $i$, accumulated over all possible paths of all lengths. The relation between the structural and the functional connectivities is thus translated to understanding how the state of one node propagates to all others along the intricate organisation of paths within the complex network.
Previous work in this direction has shown that the capacity of neural random networks to display oscillatory behaviour depends on the distribution of cycles (re-entrant paths)~\cite{Graben_Cycles_2008}, reflected by a sudden change of the network's topology when super-cycles are formed from merging of isolated loops.

From its series expansion we realise that the linear Gaussian diffusion model assumes that paths of any length are equally influential. The series only converge when the coupling is small enough such that the powers of $g^l$ decrease faster than the growth in the number of paths of length $l$ represented by $A^l$. In neural systems this scenario is rather unrealistic since information fed into the system decays rapidly due to the stochasticity of synaptic transmission and to the interaction between excitatory and inhibitory neurones in local circuits. That is, information does not perpetually propagate along the network and signals attenuate over longer processing paths. Empirical evidence from resting-state functional magnetic resonance has shown that, in general, the functional connections between regions with direct structural connections are stronger, but significant functional connections can also occur between regions without a direct connection~\cite{Goni_Communication_2014}.  While direct structural connections seem to play a major causal role in shaping the resting-state functional connectivity, the flow of information over alternative processing paths cannot be neglected. It is thus more natural to assume that shorter processing paths are more relevant than longer ones. Mathematically, the general problem is to find a set of coefficients $\{ c_l \}$ for which the series $\sum_{l=0}^\infty c_l \, A^l$ converge for any adjacency matrix and coupling strength. Although the solution to this problem is not unique, a satisfactory solution is motivated by the measure of communicability in networks~\cite{Estrada_Communicability_2008, Estrada_Review_2012}. Communicability is a generalisation of the path-length on graphs to consider a general flow of information that favours short paths over longer paths without ignoring them. There is also indications that the communicability is the Green's function of the network dynamics in case of diffusion processes, that is, the solution for the propagation of a single, infinitesimal perturbation.

The communicability between two nodes $i$ and $j$ is defined as the exponential of the adjacency matrix $\left( e^{A} \right)_{ij}$. This can itself be decomposed into a series with coefficients $c_l = 1 \, / \, l \,!$ and hence, we re-define the influence matrix $Q$ in Eq.~(\ref{eq:Qmatrix}) as:
%%%%%%%%
\begin{equation}
Q ^{exp} = \sum_{l=0}^\infty \frac{g^l A^l}{l \, !} \, = \, \mathbf{1} + gA + \frac{g^2 A^2}{2!} + \frac{g^3 A^3}{3!} + \cdots \, = \, e^{gA}.
\label{eq:Qexp}
\end{equation}
%%%%%%%%

From a physical point of view this represents the diffusion of local perturbations along the network with nonlinear (faster) decay for longer paths~\cite{Estrada_Review_2012} and it has the advantage of being free of the divergence problem of the linear Gaussian propagator. In a network nodes interact only locally with their direct neighbours, however, local perturbations propagate and can be ``sensed'' by other nodes giving rise to correlations also between distant nodes. The intensity of that correlation is thus determined by two critical factors: ($i$) the structure of the paths along which the perturbation propagates and ($ii$) the attenuation that the perturbation experiences along the way. When $g$ is weak, perturbations quickly decay giving rise to local correlations only around the perturbed node. As $g$ grows perturbations propagate deeper inside the network giving raise to stronger correlations.

Following the argumentation above, we will compute the covariance matrices as in Eq.~(\ref{eq:COV}) but replacing the propagator kernel $Q$ by $Q^{exp}$ in Eq.~(\ref{eq:Qexp}). Such a modification still keeps the simplicity and elegance of the original linear Gaussian process, but captures the physically plausible effects of convergence in the dynamical process. As shown in the Supplementary Information the results obtained with this simple mapping are consistent with those obtained after simulations of the networks using widely applied models for generic oscillatory and neural dynamics, e.g. Kuramoto oscillators and Neural-Masses.

A final note, because the coupling required to reach global synchrony depends on the size and the density of the network, the interesting range of $g$ at which the transition happens is different in every case. For convenience and for illustrative reasons we normalise the adjacency matrix by its largest real eigenvalue $\lambda_{max}$ before the calculation of $Q^{exp}$. We observed that this normalisation aligns the transition to synchrony for most networks to happen in the range $g \in [0,10)$.

%#########################################################################################
%\section*{Acknowledgements}
\acknowledgements{
The authors thank Mario Senden for discussions and early revision of the manuscript. This work has been supported by (GZL) the European Union Seventh Framework Programme FP7/2007-2013 under grant agreement number PIEF-GA-2012-331800,  the German Federal Ministry of Education and Research (Bernstein Center II, grant no. 01GQ1001A), and the European Union's Horizon 2020 research and innovation programme under grant agreement No. 720270 (HBP SGA1). YC and CSZ were supported by the Hong Kong Baptist University (HKBU) Strategic Development Fund, the Hong Kong Research Grant Council (GRF12302914), HKBU FRG2/14-15/025 and the National Natural Science Foundation of China (No. 11275027). GD is supported by the European Research Council Advanced Grant: DYSTRUCTURE (295129) and by the Spanish Research Project PSI2013-42091-P; and MLK by the European Research Council Consolidator Grant: CAREGIVING (615539).
}

%#########################################################################################

%\bibliographystyle{unsrt}
%\bibliography{Biblio_FComplexity}
%\newpage

%##########################################################################################
%############## MERGE THE SUPPLEMENTARY MATERIAL  ######################################
%\pagebreak
\clearpage
\widetext
%\clearpage

\setcounter{equation}{0}
\setcounter{figure}{0}
\setcounter{table}{0}
\setcounter{page}{1}
\setcounter{section}{0}
\makeatletter
\renewcommand{\theequation}{S\arabic{equation}}
\renewcommand{\thefigure}{S\arabic{figure}}
\renewcommand{\bibnumfmt}[1]{[S#1]}
\renewcommand{\citenumfont}[1]{S#1}

%\doublespacing
\linespread{1.5}

\begin{center}
\textbf{\large \emph{Supplementary material for:} \\ ``Functional complexity emerging from anatomical constraints in the brain: the significance of network modularity and rich-clubs.'' by }

\vspace{0.5cm}
Gorka Zamora-L\'opez, Yuhan Chen, Gustavo Deco, Morten L. Kringelbach, and Changsong Zhou

\end{center}
%%%%%%%%%% Merge with supplemental materials %%%%%%%%%%
%%%%%%%%%% Prefix a "S" to all equations, figures, tables and reset the counter %%%%%%%%%%
%%%%%%%%%% Prefix a "S" to all equations, figures, tables and reset the counter %%%%%%%%%%

\vspace{0.5cm}

\begin{quote}
\noindent Supporting information for the main article. 
In the following pages we extend the information provided in the main text. First, to better illustrate the measure of (spatial) functional complexity we show the correlation matrices for several networks along the transition to global synchrony. Second, we compare four alternative measures of functional complexity. We show that our choice, based on the integral (or area) between the observed distribution and the uniform distribution is superior to other options. Third, we revisit the `neural complexity' measure defined by Tononi, Sporns \& Edelman (1994) to emphasise its limitations and to compare it with our proposed measure. Fourth, we compare our exponential mapping to simulations of generic dynamical models. Finally, the rich-clubs of the neural networks studied in the paper are shown. This has been extensively reported in the literature before and we include them here only for completeness. Also, we add the rich-club analysis for the new hierarchical and modular network model and for the Ravasz-Barab\'asi model. 
\end{quote}

\vspace{1cm}
%#########################################################################################
\section{Evolution of correlation matrices with coupling}

\begin{figure*}[!ht]
\begin{center}
\includegraphics[width=6.83in]{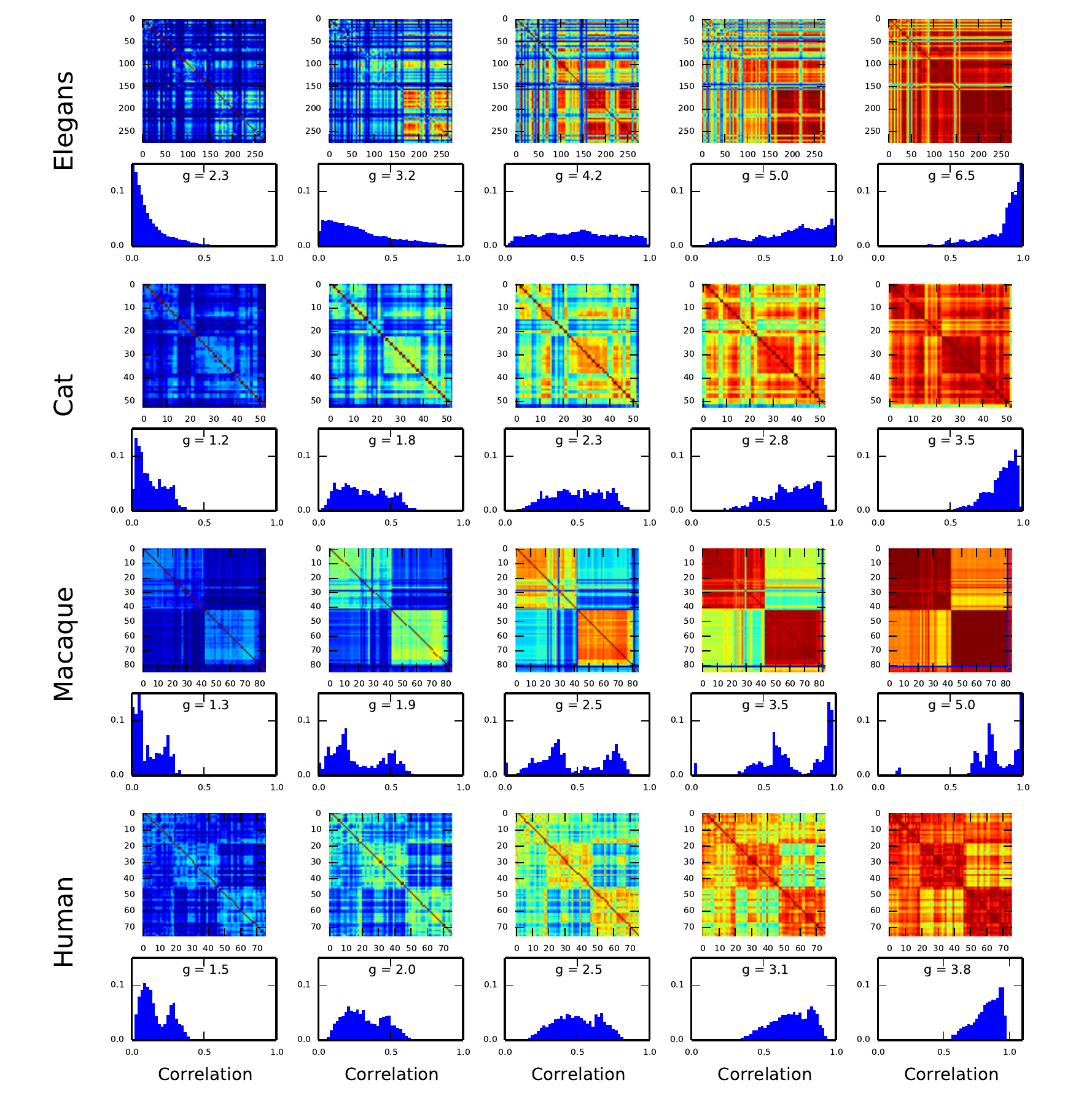}
\end{center}
\caption{{\bf Evolution of correlation matrices with increasing coupling strength $g$ for the neural connectomes.} All  matrices are adjusted to the same limits with blue corresponding to $r_{ij} = 0$ and red to $r_{ij} = 1$.
} %end of caption
\label{fig:RmatricesReal}
\end{figure*}

\begin{figure*}[!ht]
\begin{center}
\includegraphics[width=6.83in]{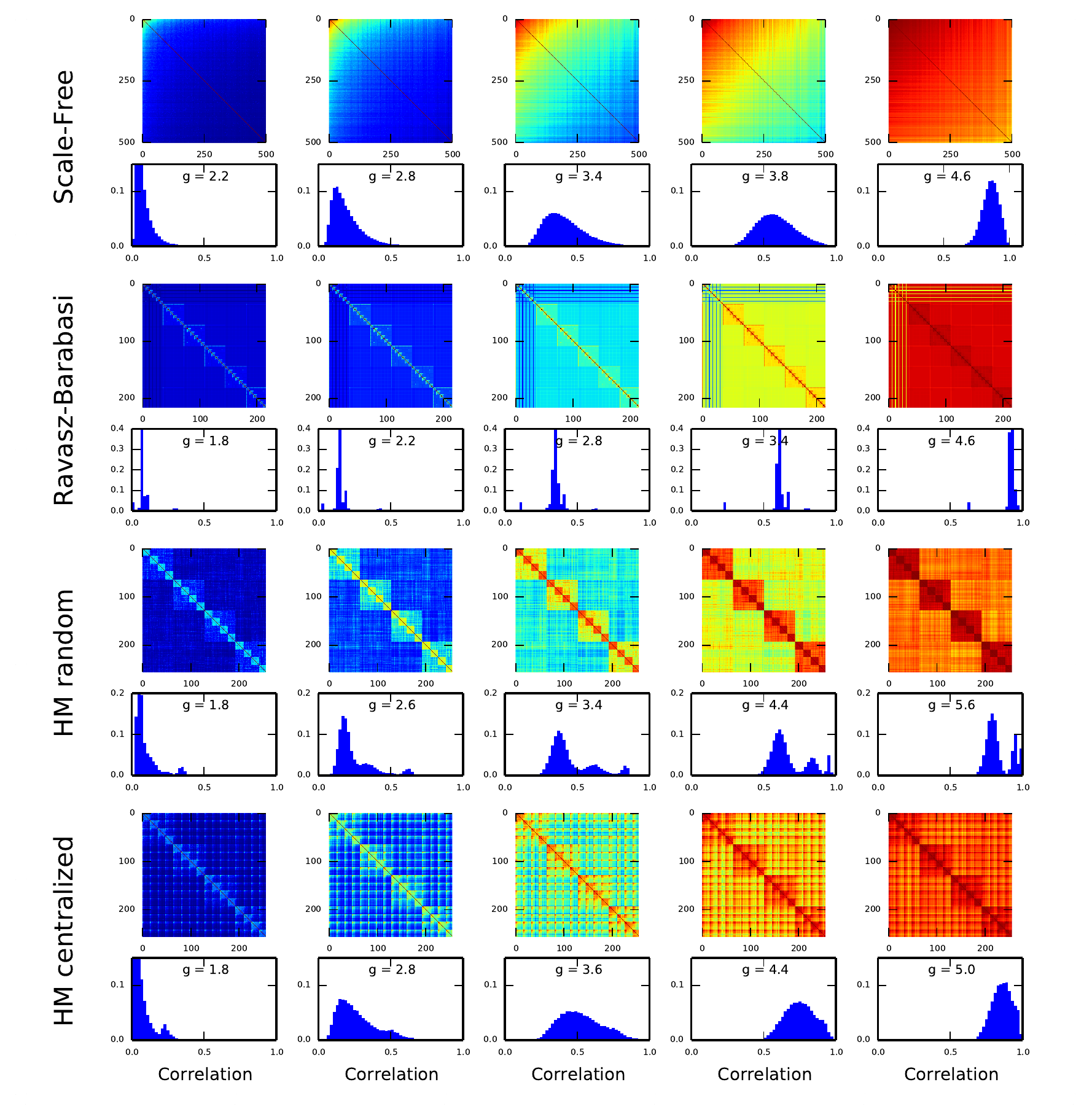}
\end{center}
\caption{{\bf Evolution of correlation matrices with increasing coupling strength $g$ for synthetic network models.} All matrices are adjusted to the same limits with blue corresponding to $r_{ij} = 0$ and red to $r_{ij} = 1$. All results are for one sample of the network models instead of ensemble averages.
} %end of caption
\label{fig:RmatricesModels}
\end{figure*}

In this paper we have introduced a measure of functional complexity which is based on the variability of pair-wise cross-correlations. To better understand the measure we show in Fig.~\ref{fig:RmatricesReal} sample correlation matrices for the neural / brain connectomes analysed in the main text. In Fig.~\ref{fig:RmatricesModels} we show sample correlation matrices for the synthetic network models. At weak coupling $g$ the distribution $p(r_{ij})$ is a narrow distribution with values near $r_{ij} = 0$. When coupling is strong, $p(r_{ij})$ becomes another narrow distribution but approaching $r_{ij} \to 1$. Complex behaviour emerges in the intermediate values of the coupling, when partial coalitions between the nodes happen; reflected by a broadening of the distribution. In the Ravasz-Barab\'asi network model $p(r_{ij})$ is always a narrow peak evidencing its lack of complex dynamics.

\newpage

\newpage

\newpage
%#########################################################################################
\section{Measures of functional complexity}

We characterise the functional complexity as the broadness of the distribution $p(r_{ij})$ of cross-correlation values $r_{ij}$. There are different manners to quantify the broadness of a distribution so here we compare the results provided by four options. The first measure is the entropy $H(p)$ of the distribution. This approach was considered in Ref.~\cite{Zhao_Complexity_2010, Zhao_Competition_2011}. Given that the distribution is evaluated with $m$ bins, the entropy of $p$ is:
%%%%%%%%
\begin{equation}
C^{ent}(p) = H(p) = - \frac{1}{C_m} \sum_{\mu=1}^m p_\mu \log p_\mu,
\label{eq:Centropy}
\end{equation}
%%%%%%%
where the normalisation constant $C_m = \log m$ is the entropy of the uniform distribution $\bar{p}$. Another option to evaluate the broadness of a distribution is to consider its variance $Var(p)$:
%%%%%%%%
\begin{equation}
C^{var}(p) = \; \frac{1}{C_m} Var(p),
\label{eq:Cvariance}
\end{equation}
%%%%%%%
where the normalisation constant $C_m = \frac{m-1}{m^2}$ is the variance of the uniform distribution $\bar{p}$. This guarantees that the measure is bounded between 0 and 1. Finally we quantify the uniformity of the distribution $p$ by directly comparing how far is the curve traced by $p$ with the curve of the uniform distribution $\bar{p}$. Therefore we consider two more choices. The first option is to compute the total Euclidean distance between the two curves. Given again that the distribution is evaluated using $m$ bins $p$ and $\bar{p}$ can be considered as two vectors in the $m$-dimensional space. Complexity is thus defined as:
%%%%%%%%
\begin{equation}
C^{euc}(p) = \; 1 \; - \; \frac{1}{C_m} \parallel p - \bar{p} \parallel, 
\label{eq:Ceuclidean}
\end{equation}
%%%%%%%
where $\parallel \cdot \parallel$ denotes the euclidean norm and $C_m = \sqrt{ \frac{m}{m-1} }$ is the distance between the Dirac-$\delta$ vector and the vector formed by the uniform distribution. Last, we define complexity as the integral (the area) between the two curves described by $p$ and $\bar{p}$.
%%%%%%%%
\begin{equation}
C^{int} = \; 1 \; - \; \frac{1}{C_m} \: \sum_{\mu=1}^m \left| \, p_\mu - \frac{1}{m}  \right|, 
\label{eq:Cintegral}
\end{equation}
%%%%%%%
where $| \cdot |$ means the absolute value and $C_m = 2 \, \frac{m-1}{m}$ is the integral between the uniform and the Dirac-$\delta$ distribution. 

\begin{figure*}[!ht]
\begin{center}
\includegraphics[width=4.6in]{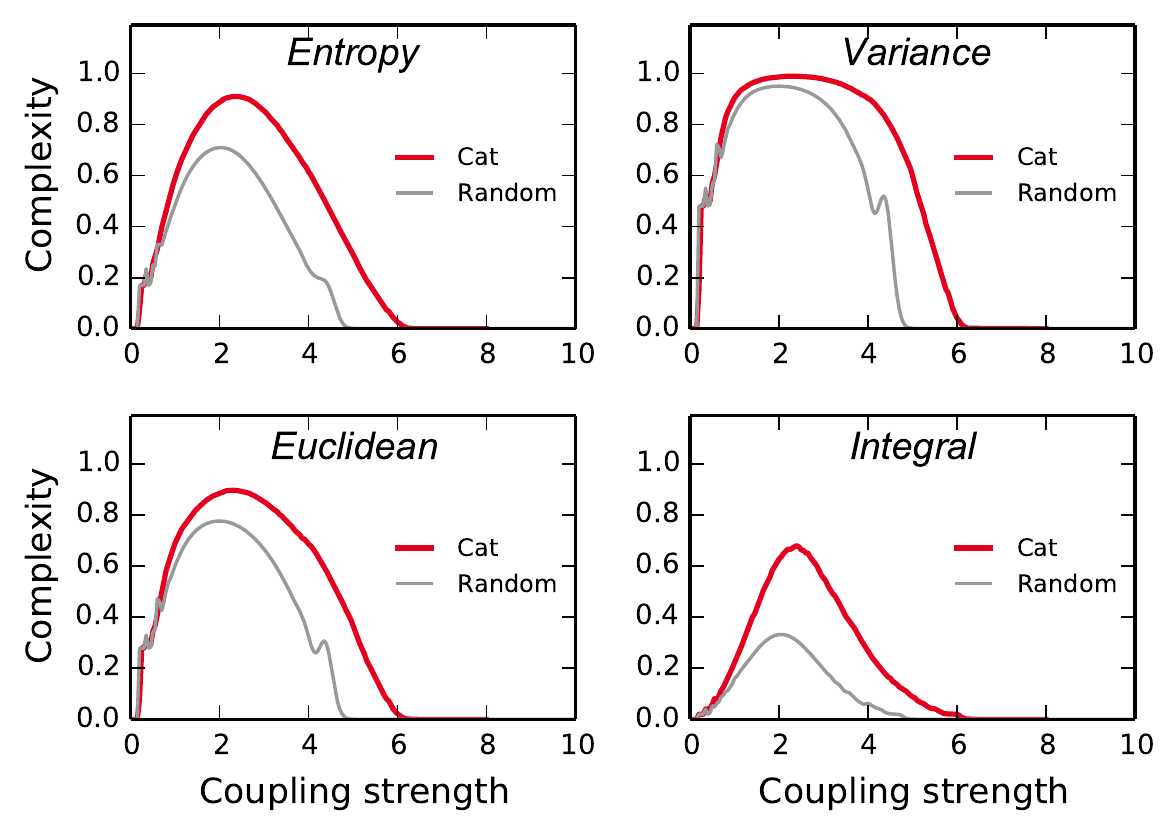}
\end{center}
\caption{{\bf Comparison of complexity measures.} Evolution of complexity as coupling increases for the cortico-cortical network of the cat and average of equivalent random graphs (200 realisations) quantified by four candidate measures of functional complexity. Entropy, variance and euclidean distance-based measures tend to overestimate the functional complexity of the networks, specially the complexity of random graphs. The integral-based measure discriminates best between the real and the random networks.
}
\label{fig:CompareComplexity1}
\end{figure*}

In order to compare the four measures we take the cortico-cortical network of the cat as an example and repeat the calculations in Fig.~2 of the main text. First we estimate the cross-correlation matrices of the network for increasing $g$ using the exponential mapping. Then we apply the four different measures of complexity to the correlation matrices and plot the results in Fig.~\ref{fig:CompareComplexity1}. For completeness we include also the evolution of complexity for equivalent random graphs of the same size and number of links as the network of the cat. As seen, entropy, variance and euclidean distance-based measures tend to overestimate the functional complexity of the network. It is particularly suspicious the large complexity these measures assign to the random graphs. In terms of discriminative power between network topologies we see that the complexity of the random graphs follow closely the complexity of the real network in the first three cases. The ratios between the peak complexity of the real network and of the equivalent random graphs $r = \frac{C_{max}(cat)}{C_{max}(random)}$ are: $r^{ent} = 1.296$, $r^{var} = 1.046$, $r^{euc} = 1.190$ and $r^{int} = 2.048$. The integral-based measure of complexity is, by far, the best measure among the four to discriminate between network topologies.

\begin{figure*}[!ht]
\begin{center}
\includegraphics[width=4.6in]{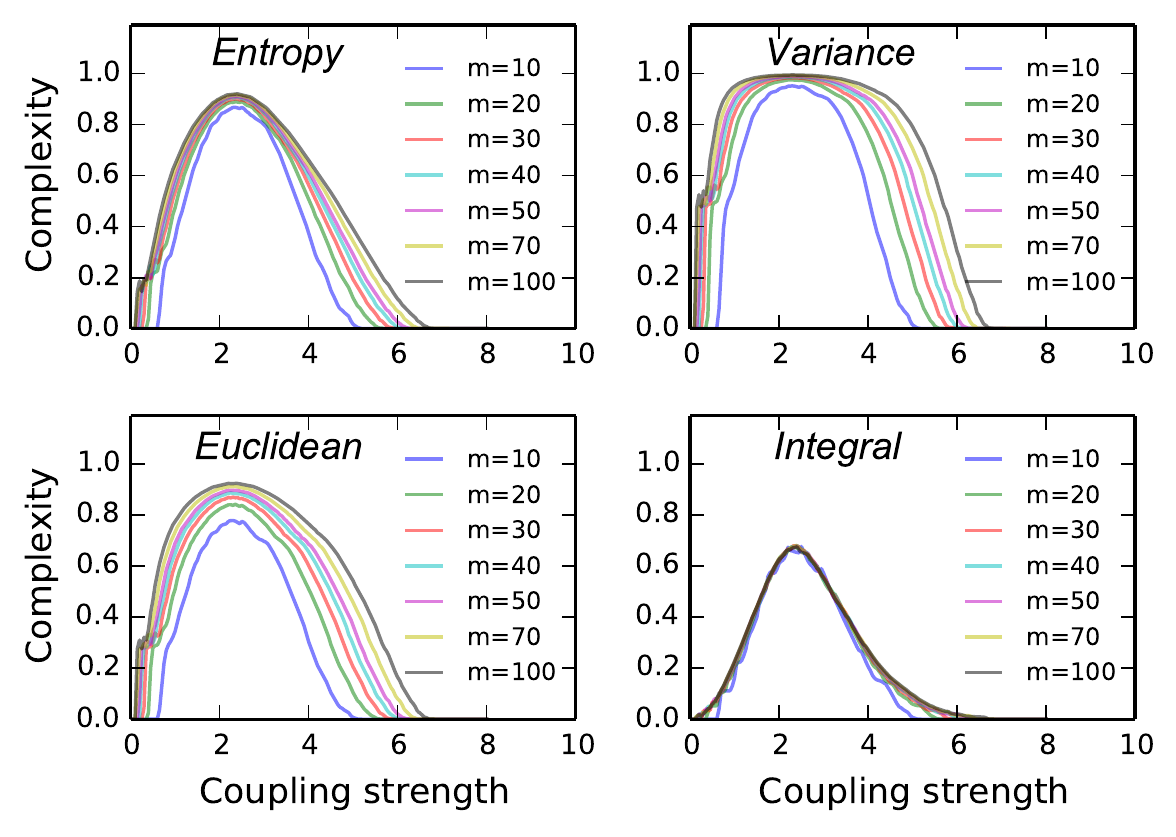}
\end{center}
\caption{{\bf Robustness of complexity measures.} The evolution of the complexity as coupling increases for the cortico-cortical network of the cat as quantified by four candidate measures of functional complexity. Before computing complexity, the distribution $p(r_{ij})$ was evaluated using different numbers of bins $m$. The integral-based measure is the only robust measure of the four.
}
\label{fig:CompareComplexity2}
\end{figure*}

We now investigate their robustness against arbitrary variation in the number of bins used to estimate the distribution $p$. In Fig.~\ref{fig:CompareComplexity2} we plot again the evolution of the complexities for the corticocortical network of the cat as the coupling strength increases. The only difference now is that before computing complexity, the distribution $p(r_{ij})$ is estimated using a different number of bins $m$ to cover the range $r_{ij} \in [0, 1 ]$. The only measure that returns robust results is the integral-based measure of complexity. The results of the other three measures clearly depend on the number of bins.

\newpage

\newpage
%#########################################################################################
\section{Limitations of the \emph{Neural Complexity} measure}

Tononi, Sporns \& Edelman (1994) introduced a measure of complexity, named as `neural complexity', intended to quantify the balanced coexistence of both local and global collective coherent behaviour in a dynamical network. The main idea was that the measure would become largest for network which can balance between segregation and integration. Functional segregation is regarded as the relative statistical independence between groups of elements of the system, and functional integration as the statistical dependence between the groups. A system shall be complex when it contains dynamical clusters that are weakly correlated between them.

Given a multivariate dynamical system (or network) $X$ consisting of $N$ coupled components (nodes), neural complexity was defined as the sum of the average mutual information of all possible bipartitions in the network, for bipartitions of size $k = 1, 2, \ldots, N/2$:
%%%%%%%%
\begin{equation}
C^N(X) = \sum_{k=1}^{N/2} \left< MI(X_j^k ; \tilde{X}_j^k) \right>.
\label{eq:TScomplexity}
\end{equation}
%%%%%%%%
Here $MI$ stands for mutual information and $\{X^k, \tilde{X}^k \}$ is a bipartition of the network into two complementary subsets of sizes $k$ and $N-k$.

The mutual information between two subsets can be computed in terms of the integration $I$, also defined in~Ref.\cite{Tononi_Complexity_1994}. Integration is a generalisation of mutual information for more than two variables: $I(X) = \sum_{i=1}^N H(x_i) - H(X)$. Here $H(x_i)$ is the entropy of each variable (node) and $H(X)$ is the joint entropy of the coupled system as a whole. The mutual information between two complementary subsets can be rewritten as:
%%%%%%%%
\begin{equation}
MI(X^k ; \tilde{X}^k) = I(X) - I(X^k) - I(\tilde{X}^k).
\label{eq:MI}
\end{equation}
%%%%%%%%
In real applications, measuring the joint distribution of multivariate time-series can be unfeasible because it requires large amounts of data to be available. This problem can be avoided by estimating integration $I(X)$ out of the pairwise cross-correlation matrix $R(X)$ of the system as: $I(X) = - \frac{1}{2} \log \left( |R(X)| \right)$ where $|R|$ stand for the determinant of the correlation matrix. Substituting in Eq.~(\ref{eq:MI}), the mutual information for a given bipartition becomes:
%%%%%%%%
\begin{eqnarray}
MI(X^k ; \tilde{X}^k) & = & \frac{1}{2} \Big(  \log |R(X^k)| + \log |R(\tilde{X}^k)| - \log |R(X)|  \Big) \\
                            & = & \frac{1}{2} \log \left[ \frac{|R(X^k)| \; |R(\tilde{X}^k)|}{|R(X)|} \right]
\label{eq:MIR}
\end{eqnarray}
%%%%%%%
where $R(X^k)$ and $R(\tilde{X}^k)$ are two sub-matrices of $R(X)$ taking only the nodes in the subsets $X^k$ and $\tilde{X}^k$ respectively.

\begin{figure*}[!ht]
\begin{center}
\includegraphics[width=4.86in]{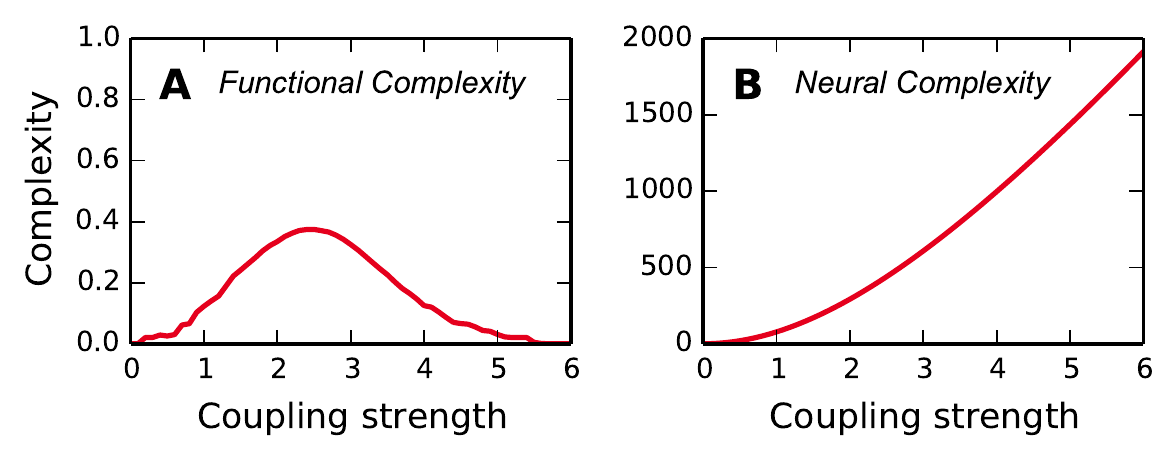}
\end{center}
\caption{{\bf Comparison with `neural complexity' measures.} The evolution of the complexity for one random graph of $N = 100$ nodes and density $\rho = 0.2$ as the coupling strength increases. For each value of $g$ the correlation matrix $R$ of the network is estimated with the exponential mapping and the two complexities are then calculated out of the same $R$. (A) Shows the result for functional complexity (Eq.~(\ref{eq:Cintegral})) and (B) the result for neural complexity $C^N(X)$.
}
\label{fig:CompareComplexity3}
\end{figure*}

Neural Complexity suffers from few limitations. On the one hand, it requires to compute the mutual information between all possible bipartitions. For a network of size $N$ there are $\sum_{k=0}^{N/2} \frac{N!}{k! \, (N-k)!}$ such bipartitions making it computationally feasible only for very small networks. The problem can be partly overcome by estimating the result from a smaller random sample of bipartitions. More critically, the measure takes its largest value when the system is globally synchronised, diverging to infinity. When the nodes are uncoupled $C^N(X) = 0$ as it is expected. As the coupling strength of the links increases $C^N(X)$ grows monotonically. The problem is that when all values of the correlation matrix are $r_{ij} = 1$, the mutual information between any bipartition $MI(X^k ; X - X^k)$ in Eq.~(\ref{eq:TScomplexity}) becomes infinite. To demonstrate this, imagine that the network is almost synchronised; there is a small number $0 < \delta \ll 1$ such that $r_{ii} = 1$ and $r_{ij} = 1 - \delta$ for all $i \neq j$:
%%%%%%%%
\begin{equation} 
COR(X) = \left( 
	\begin{array}{cccc}  
	1 & 1-\delta & 1-\delta & \cdots  \\ 
	1-\delta & 1 & 1-\delta  & \cdots \\ 
	1-\delta & 1-\delta & 1 & \cdots \\
	\vdots & \vdots & \vdots & \ddots \\
	\end{array}
\right) _{(N \times N)}.
\end{equation}
%%%%%%%%
In this case the determinant of $R$ can be easily expressed and integration reduces to 
%%%%%%%%
\begin{equation}
I(x) = -0.5 \log \left( \, |COR(x)| \, \right) = - \log \left( N\delta^{N-1}-(N-1)\delta^N \right).
\end{equation}
%%%%%%%
Substituting in Eq.~(\ref{eq:MIR}) we obtain that the mutual information for a given bipartition is: 
%%%%%%%%
\begin{equation}
MI(X^k_j; \tilde{X}^k_j) = 0.5 \, \log \left[ \, \frac{ \left[ \, (k \, (1-\delta) + \delta \, \right] \; \left[ \, (N-k)(1-\delta) + \delta \, \right]}
{\delta \, \left[ \,N (1 - \delta) + \delta \, \right]} \; \right].
\end{equation}
%%%%%%%
When $\delta$ is very small, $0 < \delta \ll 1$, this expression can be approximated by:
%%%%%%%%
\begin{equation}
MI(X^k_j; \tilde{X}^k_j) = 0.5 \, \log \left[ \, \frac{k \, (N-k)} {\delta \,N} \, \right]
\end{equation}
%%%%%%%
which diverges to infinity as $\delta \to 0$. Thus, neural complexity $C^N(X)$ becomes infinity in the globally synchronised state what is contradictory with the intention of the measure. At the globally synchronised state there is no segregation and hence there is no optimal balance between segregation and integration. Figure~\ref{fig:CompareComplexity3} shows the numerical comparison between functional complexity and neural complexity applied to the same random graph of $N = 100$ nodes. As seen, $C^N(X)$ monotonically increases with coupling. Our measure, on the contrary, successfully vanishes again at strong $g$.

%#########################################################################################
\section{Comparison of the exponential mapping with dynamical models}

The principal goal of this paper is to develop an exploratory \emph{proof of concept} on how modular and hierarchical network organisation with rich-club forming hubs enhance the complexity of the networks. For that we have studied the spatial formation of clusters and their interactions. For computational convenience we have analytically estimated the time-averaged correlation matrices with a mapping that accounts for the nonlinear decay of signals over longer paths. Also, the only parameter of the mapping is the coupling strength between nodes and allows a direct comparison between the structural and the functional connectivities. When the local dynamics depend on several parameters it is not always possible to discern whether the observed collective dynamics are shaped by the network's topology or are triggered by the local parameters~\cite{Schmidt_2010}.

In the following we show that the exponential mapping we have introduced represents a plausible \emph{effective approximation} of the correlations between nodes in a network. For that we compare the results obtained for the corticocortical network of cats with the results achieved by simulating the network with two generic models of oscillatory and neural activity, the Kuramoto model and the neural-mass model. For completeness we also include results for the Gaussian noise diffusion model, Eq.~(3) in the main text.

{\bf The Kuramoto model} is a norm form of interacting self-sustained oscillators. It represents the mean-field dynamics of a population of weakly coupled and nearly identical limit cycle oscillators~\cite{Kuramoto_Book_1984}. The phase of each oscillator $\theta_i$ is described as:
%%%%%%%%
\begin{equation}
\dot{\theta}_i \; = \; \omega_i + \;g \, \sum_{j=1}^N A_{ij} \sin(\theta_j - \theta_i),
\end{equation}
%%%%%%%
where $\omega_i$ are the natural frequencies and $g$ is the coupling strength. For the simulations we set the natural frequencies at random from a normal distribution with mean $\bar{\omega} = 1$ and variance $0.0025$. This guarantees that the frequency of the fastest node is less than twice the frequency of the slowest one. Initial conditions $\theta_i(0)$ were chosen uniformly at random from values in the range $[-\pi,\pi]$. Cross-correlation between regions was calculated out of the sinusoidal signals, $x_i(t) = \sin \left( 2 \pi \, \theta_i(t) \right)$. Functional complexity and the mean correlations were calculated out of the average correlation matrix after 200 realisations.

{\bf The neural-mass model} was designed to reproduce the macroscopic rhythmic activity of cortical regions similar to that observed with EEG or MEG~\cite{Jansen_NeuralMass_1995, Wendling_MassModel_2000, David_NeuralMass_2003}. The model for one cortical region consists of three interconnected neuronal subpopulations: two of excitatory neurones and one of inhibitory neurones. The average membrane potential of each subpopulation is represented as a critically dumped harmonic oscillator of the form $\ddot{v} = - 2a \, \dot{v} - a^2 \, v$, with the inhibitory population having slower decay rate. The excitatory interneurons receive a constant noisy input of the form $p(t) = p_0 + \xi(t)$ where $\xi(t)$ is a Gaussian white noise. The baseline $p_0$ controls for the frequency at which the mass model. Denoting the mean membrane potential of the excitatory, pyramidal and inhibitory subpopulations of region $i$ as $E_i$, $P_i$ and $I_i$ the coupled system is written as:
%%%%%%%%%
\begin{eqnarray}
\ddot{E_i} & = & a A \left[ C_2 \, f(C_1 P_i) \, + \, p(t) \, + \, g \, \sum_{j=1}^N A_{ij} f(E_j - I_j) \right] \, - \, 2a \dot{E_i} \, - \, a^2 E_i ,\\
\ddot{I_i} & = & b B \, C_4 \, f(C_3 P_i)  \; - \, 2b \dot{I_i} \, - \, b^2 I_i, \\
\ddot{P_i} & = & a A \, f(E_i - I_i)  \quad - \, 2a \dot{P_i} \, - \, a^2 P_i,
\end{eqnarray}
%%%%%%%%
where $A$ and $B$ represent the average synaptic gains, $1/a$ and $1/b$ the average dendritic-membrane time constants. $C_1$ and $C_2$, $C_3$ and $C_4$ are the average number of synaptic contacts between subpopulations, for the excitatory and inhibitory synapses, respectively. A static nonlinear sigmoid function $f(v)=2e_0/(1+e^{r(v_0-v)})$ converts the average membrane potential into an average pulse density of action potentials. Here $v_0$ is the postsynaptic potential corresponding to a firing rate of $e_0$, and $r$ is the steepness of the activation. Together with the external noisy input, the input from other regions are fed into the excitatory subpopulation of interneurons. We have considered the same model parameters as in references~\cite{Wendling_MassModel_2000, Zhou_NJP_2007} to generate oscillations in the $\alpha$ band with the exception of two parameters. We set $e_0 = 3.0 \, s^{-1}$ and $p_0 = 200 \, mV$ to stabilise the oscillations.

\begin{figure*}[!ht]
\begin{center}
\includegraphics[width=6.83in]{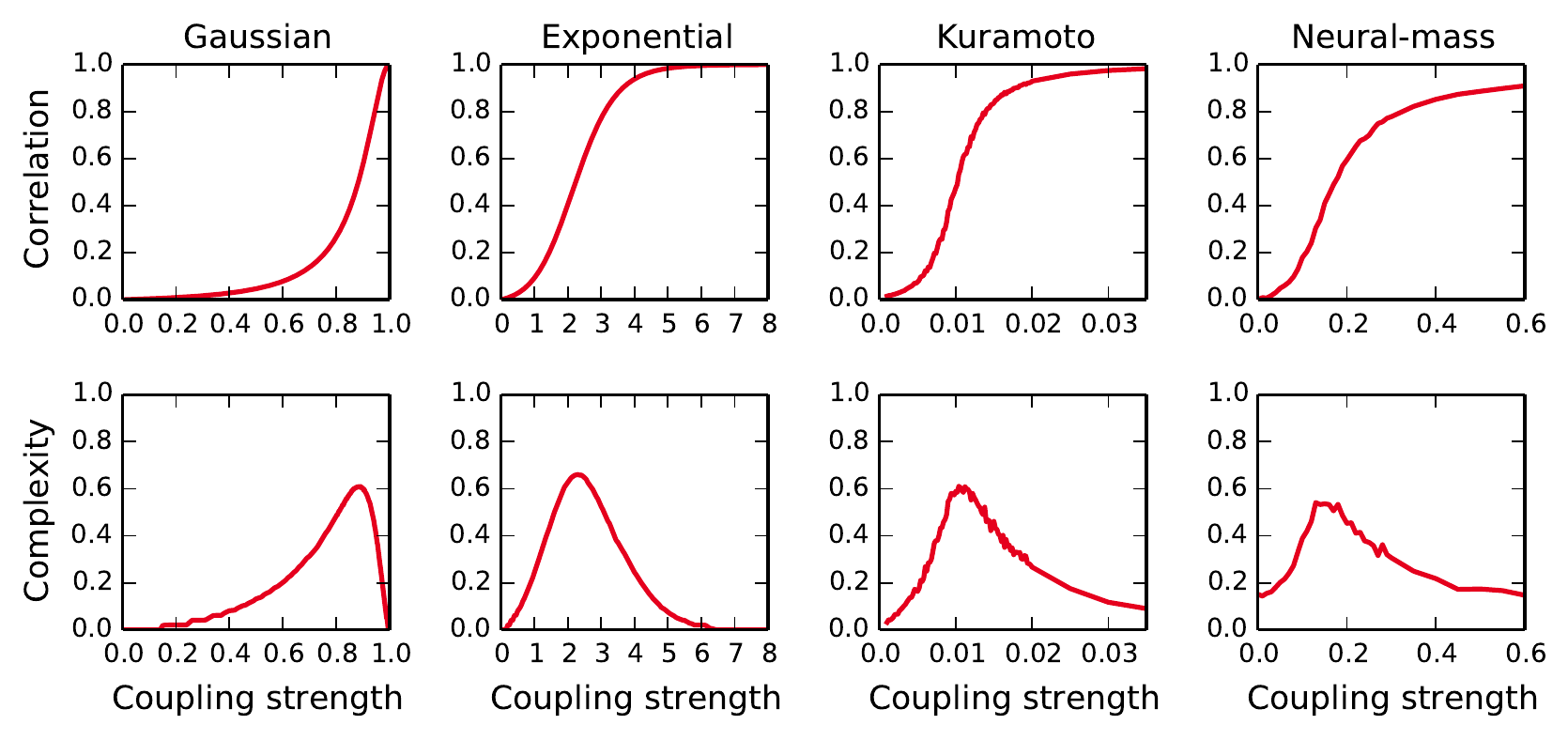}
\end{center}
\caption{{\bf Functional complexity under different dynamical models.} The evolution of mean correlation and of functional complexity in the corticocortical network of cats are compared when the network is simulated using four different dynamical models: ($i$) a linear diffusion process of Gaussian noise, ($ii$) a nonlinear diffusion (exponential mapping) of Gaussian noise, ($iii$) a network of self-sustained Kuramoto oscillators and ($iv$) a network of neural-mass models.
} %end of caption
\label{fig:CompareModels}
\end{figure*}

We run the simulations on the corticocortical network of the cat with neural-mass models using Euler's integration method with step $dt = 0.01 \, ms$ for 20 seconds. Since the regions are assumed identical (they all have same parameters) choosing uniform random initial conditions introduces spurious correlations into the network. Therefore the first 5 seconds of each run were performed uncoupled ($g = 0$) with a strong Gaussian noise with variance $10\%$ the amplitude of $p_0$. Afterwards, the coupling was switched on and the noise level reduced to 2\% of $p_0$. 100 realisations were simulated for every value of $g$. Functional complexity and mean correlations were computed out of the $z$-Fisher corrected average correlation matrix.

In Supplementary Fig.~\ref{fig:CompareModels} we show the mean correlation and the functional complexity of the corticocortical network of cats when the system dynamics are simulated with four different models: ($i$) a linear diffusion of Gaussian noise, Eq.~(3) of main text, ($ii$) an exponentially decaying diffusion process, ($iii$) coupled Kuramoto oscillators and ($iv$) coupled neural mass models. The qualitative behaviour of the network is very robust: in all the four cases a transition to global synchrony is observed as coupling increases. The neural mass model does not fully achieve global synchrony because of the external noise. The functional complexity shows its characteristic shape in the four cases, vanishing in the extremes with a maxima in between. The four models achieve similar peak values of complexity between 0.5 and 0.6. For the linear Gaussian model we observe that the interesting regime happens at rather high values of $g$, near the critical coupling at which the system diverges. These results corroborate the plausibility of our exponential mapping as a proxy of collective dynamics in a network of generic oscillators.

%#########################################################################################
\section{Rich-club of neural and synthetic networks}

A complex network is said to have a rich-club when the nodes with largest degree are densely interconnected. To quantify this behaviour Zhou and Mondrag\'on introduced the measure $k$-density, $\Phi(k')$, which is the density of links in the subnetwork composed by the nodes with degree $k > k'$~\cite{Zhou_RichClub_2004}. In other words, $\Phi(k')$ is the ratio between number of links $L'$ contained in the subnetwork composed by the nodes with degree $k > k'$ and all the links possible $^1\! / _2 \; N' (N'-1)$ in that subnetwork. $N'$ is the number of nodes with $k > k'$. The factor $^1 \! / _2$ is applied for undirected networks. Formally written:
%%%%%%%%
\begin{equation}
\Phi(k') = \frac{2 L'} {N' (N'-1)}.
\label{eq:kdensity}
\end{equation}
%%%%%%%%
Now, $\Phi(k')$ can be repeatedly applied for all $k' = 0, 1, 2, \ldots, k^{max}-1$ (where $k^{max}$ is the largest degree observed in the network) and draw the resulting curve. The initial point at $k' = 0$ is the original density of links of the network. The question is thus \emph{whether for successive $k'$ the curve grows above the initial density $\Phi(0)$, whether it remains stable or whether it decreases below $\Phi(0)$}.
If $\phi(k)$ decays, then we are sure there is no rich-club in the network. If $\phi(k)$ grows, then maybe.

\begin{figure*}[!ht]
\begin{center}
\includegraphics[width=6.83in]{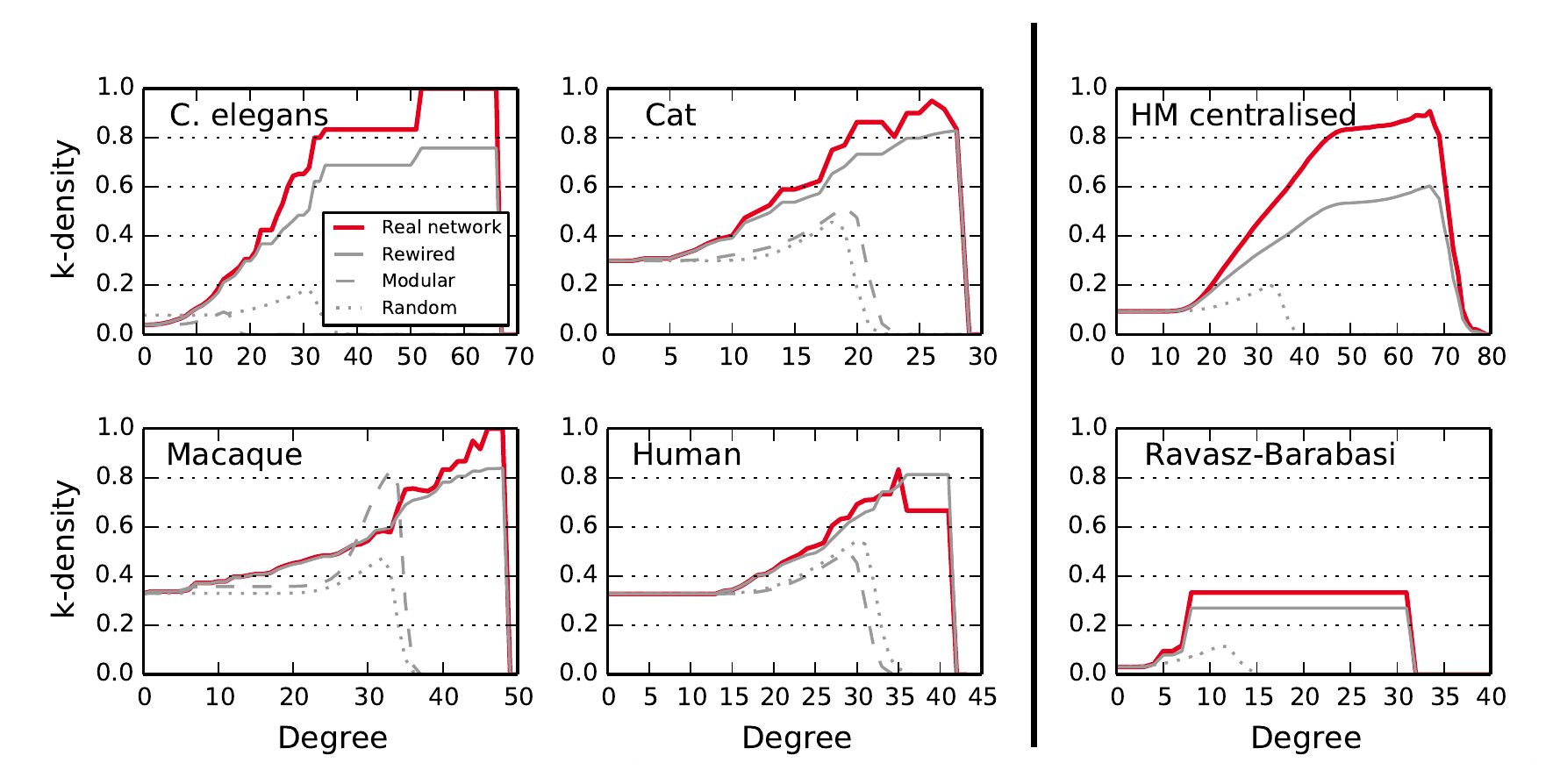}
\end{center}
\caption{{\bf $k$-density of real and model networks.} All results for rewired and random networks are the average curves for ensembles of 1000 realisations, except for the HM centralised model. Since the HM centralised model is stochastic and every realisation is different, we generated 100 networks. The black solid line is their average curve. For each of the 100 networks, 100 realisations of the rewired and of the random graphs were created.
} %end of caption
\label{fig:RichClub}
\end{figure*}

Three of the four real networks investigated are directed. Since the $k$-density is a priori defined for undirected networks, in those cases we define the degree of node $i$ as the average of its input and output degrees: $k = 0.5 \, (k^{in}_i + k^{out}_i)$. This is a reasonable approximation due to the high fraction of reciprocal connections in these networks and the large correlation between input and output degrees. For more detailed applications $k$-density can be computed ranking the nodes according to their $k^{in}$ or their $k^{out}$ separately. The results in Fig.~\ref{fig:RichClub} show how the $k$-density of the four real networks (black solid lines) monotonically ascend and reach very large densities, a clear indication of the presence of a rich-club. To identify the composition of the rich-club in each network we considered the set of hubs remaining at the degree $k'$ for which $\Phi(k') \leq 0.8$.
An ideal rich-club is a set of hubs which are all-to-all connected. In that case $k$-density reaches its maximal value $\Phi(k)=1.0$. Our choice to consider rich-clubs when $\Phi(k)=0.8$ is to set a rather conservative criteria and be sure that the rich-clubs we observe are ``close enough'' to the ideal case. Our reference is the \emph{closeness} to the ideal rich-club, instead of \emph{how expected} $\phi(k)$ is compared to rewired networks with same degree distribution. It has to be noted, however, that there might other (lower) values of $k$ at which $\Phi(k) < 0.8$ in absolute value, but for which $\phi(k)$ is more surprising in the comparison to the expected value $\Phi_{rew}(k)$ for rewired graphs with same degree distribution.

The table below summarises  their rich-club properties such as the degree for which the $k$-density becomes larger than 0.8, the number of nodes remaining at that $k'$ and the actual density, $\Phi(k')$, of the subnetwork formed by them.\\

\begin{tabular}{p{1.8cm} p{0.7cm} c c | p{8cm}}
{\bf Network} 			& $\mathbf{k'}$ & $\Phi\mathbf{(k')}$ & {\bf Size (nodes)}  & {\bf Neurones or cortical areas}  \\
\hline
\emph{C. elegans}	& 32 & 0.833 & 5 	& AVAL, AVAR, AVBL, AVBR, PVCR   \\ 
Cat 				& 23 & 0.864 & 11 	& 20a, 7, AES, EPp, 6m, 5Al, 1a, 1g, CGp, 35, 36  \\
Macaque			& 40 & 0.833 & 7 	& 46, 7a, 7b, LIPd, LIPv, MT, VIPl  \\
Human			& 35 & 0.833 & 6	&  Precuneus (L/R), FrontSup. (L/R), OccMid. (L/R)  \\
\hline
\end{tabular}

%\vspace{0.5cm}

For comparison we include also the ensemble average $k$-density curves for three null-models, the same used in the comparisons of complexity: ($i$) rewired networks conserving the degree of the nodes, ($ii$) random graphs of same size and number of links as the network, and ($iii$) random networks with the same modular structure as the original network (see Materials and Methods section in the main text). As expected, the rewired networks (dashed lines) follow closely the $k$-density of the real networks. At the largest degrees, however, the real networks still achieve the largest values. In the case of the macaque and the human tractography this relationship is the closest implying that the presence of the rich-club might be ``explained'' by their degree distribution alone. Random networks (dotted lines) and modularity preserving random networks (dash-dotted lines) also tend to increase $\Phi(k')$ although significantly slower than the real and the rewired networks. They only reach maximal densities slightly above the initial $\Phi(0)$. The early cut-off is because the degree distribution of random graphs is a Poissonian distribution with all nodes having degree comparable to the mean. The neural and brain networks, however have a broad degree distribution with largest degrees above the expectation in random graphs.

Finally, the $k$-density of the hierarchically modular network with centralised intra-modular connectivity (Centralised HM model) is shown in Fig.~\ref{fig:RichClub} to corroborate that the model gives rise to a rich-club with the parameters used for the results in Fig.~7 of the main text. The $k$-density for the Ravasz-Barab\'asi model demonstrates that the model fails to generate a rich-club despite it has a scale-free-like degree distribution.

%#########################################################################
\section*{References}
% The bibtex filename
%\bibliographystyle{unsrt}
%\bibliography{Biblio_FComplexity_SI.bib}

\end{document}